\documentclass[aps,pra,10pt,twocolumn,floatfix,nofootinbib,superscriptaddress]{revtex4-2}
\usepackage[english]{babel}
\usepackage[utf8]{inputenc}
\usepackage[normalem]{ulem}
\usepackage{graphicx}
\usepackage{amsmath,amssymb,amsfonts,physics,graphicx,xcolor,mathtools,dsfont,bm}
\usepackage[noend]{algpseudocode}
\usepackage[]{todonotes} % [disable] to remove notes
\usepackage{xfrac}
\usepackage{booktabs}
\usepackage{placeins}
\usepackage{soul}
\usepackage{tikz}
\usetikzlibrary{shapes.geometric, arrows.meta, positioning, calc, backgrounds, fit}
\usepackage{hyperref}
\hypersetup{
	colorlinks=true,
	linkcolor={red!40!black},
	citecolor={blue!60!black},
	urlcolor={blue!50!black}
	}
\graphicspath{{fig/}}
\definecolor{myred}{rgb}{0.85,0,0}
\definecolor{mygray}{rgb}{0.87, 0.87, 0.87}

\newcounter{algorithm}
\renewcommand{\thealgorithm}{\arabic{algorithm}}
\makeatletter
\newcommand{\fnum@algorithm}{Algorithm~\thealgorithm}
\def\ext@algorithm{loa}
\newenvironment{algorithm}{%
  \par\addvspace{0.5\baselineskip}%
  \begingroup
  \refstepcounter{algorithm}%
  \def\caption##1{\hrule height 0.8pt\relax\vskip 3pt%
    \noindent\textbf{Algorithm~\thealgorithm: ##1}\par\vskip 3pt\hrule\vskip 4pt}%
}{%
  \par\vskip 3pt\hrule height 0.8pt\relax\endgroup\addvspace{0.5\baselineskip}%
}
\makeatother
\let\oldbibitem\bibitem % copy bibitem into oldbibitem
\renewcommand{\bibitem}{
 \renewcommand{\doi}[1]{\texttt{\href{https://doi.org/##1}{doi:##1}}} % override \doi
 \let\bibitem\oldbibitem % restore \bibitem
 \oldbibitem % call old \bibitem
}
\begin{document}
% Force fully justified captions (squared: flush left AND right)
\makeatletter
\long\def\@makecaption#1#2{%
	\vskip\abovecaptionskip
	{\leftskip=0pt \rightskip=0pt \parfillskip=0pt plus 1fil
		\noindent #1: #2\par}%
	\vskip\belowcaptionskip}
\makeatother
\title{Quantum Community Detection for Data Analysis}
\author{Vicente P. Soloviev}
\affiliation{Fujitsu Research of Europe Ltd., Pozuelo de Alarcón, 28224 Madrid, Spain}
\email{vicente.perezsoloviev@fujitsu.com}
\author{Brian Coyle}
\affiliation{Fujitsu Research of Europe Ltd., Slough SL1 2BE, UK}
\affiliation{School of Informatics, University of Edinburgh, Edinburgh, UK}
\author{Stefano Scali}
\affiliation{Fujitsu Research of Europe Ltd., Slough SL1 2BE, UK}
\affiliation{University of Exeter, Department of Physics and Astronomy, Stocker Road,
Exeter EX4 4QL, UK}
\author{Michal Krompiec}
\affiliation{Fujitsu Research of Europe Ltd., Slough SL1 2BE, UK}
\begin{abstract}
	Community detection offers an alternative to conventional methods for data
analysis tasks such as feature selection and clustering, and detecting communities with
quantum algorithms is a promising avenue for quantum speedup. We investigate quantum
spectral community detectors, such as the modularity-matrix algorithm deteQt, in
pipelines for both tasks. We first generalise deteQt to multiple communities by
recursive bisection, and show that it recovers planted communities, and hence redundant
features, in synthetic and small datasets. We then propose projeQt, which applies a
spectral filter to the encoded modularity matrix through the quantum singular value
transformation, retaining higher-order eigenvectors that deteQt does not use. projeQt
acts on the generalised modularity matrix and extracts several communities at once,
without recursion, by $k$-means clustering of a Gram matrix built from quantum state
overlaps. On MNIST we demonstrate both pipelines, selecting informative pixels and
classifying images directly by clustering, and find them competitive with standard
community detectors.
\end{abstract}
\maketitle
\section{Introduction} \label{sec:introduction}
The goal of community detection algorithms is to partition a graph into groups of nodes
that are more densely connected to each other than to the rest of the
network~\cite{fortunato2010community,fortunato2016community,fortunato2022twenty}. Many
algorithms do this by maximising modularity~\cite{newman2004finding}, which compares
the density of edges inside the proposed groups with that expected in a random graph
with the same degrees. Maximising it exactly is NP-hard~\cite{brandes2008modularity},
so practical methods are approximate. Divisive methods progressively remove the edges
of highest betweenness, which lie between communities~\cite{girvan2002community}, and
keep the partition with the highest modularity among those
produced~\cite{newman2004finding}. Greedy methods such as the Louvain
algorithm~\cite{blondel2008}, and its refinement Leiden~\cite{traag2019louvain}, move
each node into the neighbouring community that most increases the modularity, then
merge each community into a single node and repeat on the smaller graph until the
modularity no longer increases. Community detection can also be posed as statistical
inference of a stochastic block
model~\cite{holland1983stochastic,karrer2011stochastic,peixoto2014hierarchical}, and
maximising modularity is equivalent to maximum-likelihood inference of a special case
of that model~\cite{newman2016equivalence}. In this work, we will be primarily focused
on spectral modularity methods, which divide the graph using the leading eigenvectors
of the modularity matrix~\cite{newman2006modularity,newman2006finding}.
\begin{figure*}[!t]
	\centering
	\includegraphics[width=\textwidth]{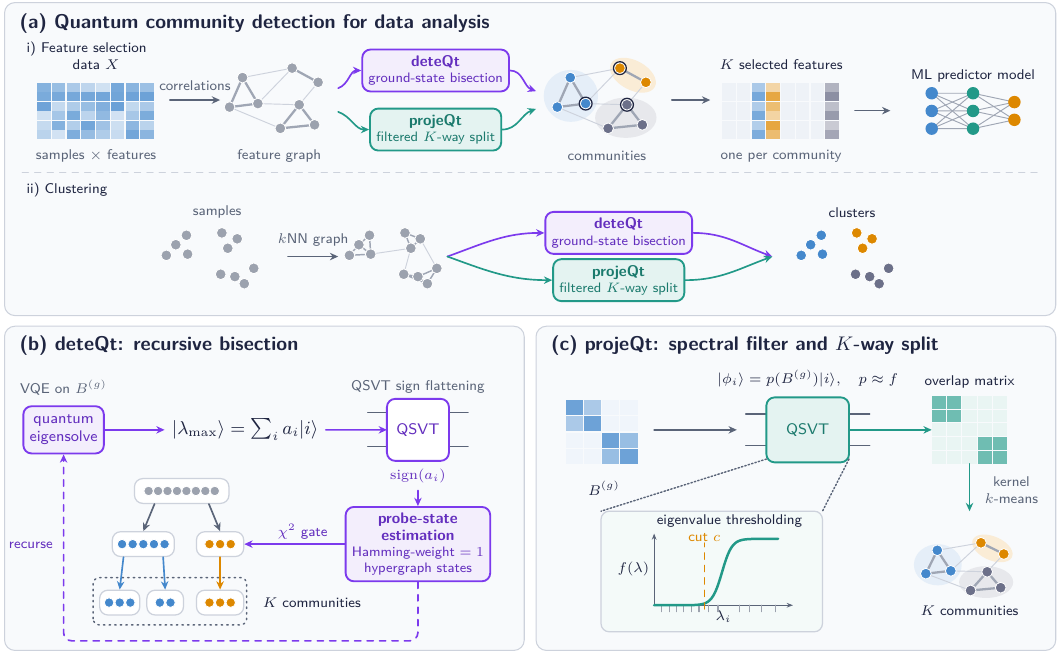}
	\caption{\textbf{Community detection for data analysis.} \textbf{(a)}~For
feature selection, correlated features form a graph community detection selects one
representative feature per community for e.g. an ML predictor. For clustering, samples form
a $k$-nearest-neighbour graph and the detected communities are the clusters directly. Community detection is done via the deteQt and projeQt quantum algorithms.
\textbf{(b)}~VQE prepares the leading eigenvector of the current group's modularity
matrix. Amplitude block encoding and QSVT apply the sign function to its amplitudes to
propose a bisection. Accepted groups are split recursively until $K$ communities are
reached. \textbf{(c)}~projeQt applies a QSVT spectral filter to the group's generalised
modularity matrix, estimates overlaps between filtered node states, and clusters them
with kernel $k$-means to form $K$ communities.}
	\label{fig:hero}
\end{figure*}
However, all of the above classical community detection methods have limitations.
Louvain, being a greedy method, finds partitions which depend on the order of visited
nodes, rather than on the value of modularity it achieves~\cite{blondel2008}, despite
running in close to linear time on sparse graphs~\cite{blondel2008,traag2015faster}.
Modularity-based methods also have a resolution limit~\cite{fortunato2007resolution}.
In a graph containing $m$ edges, take an example where we have two communities which
are connected by only a single edge, yet each community has fewer than about
$\sqrt{m/2}$ internal edges. Such cases have a higher modularity when the two
communities are merged than separated, even if each community is itself a clique. As a
result, any modularity-maximising method can merge two separate, distinct communities.
On the other hand, spectral methods can detect communities down to the detectability
threshold, the limit for any method, in graphs of large mean
degree~\cite{nadakuditi2012graph,decelle2011asymptotic}, but are expensive since they
typically require the leading eigenvector(s) at every community split. Using the power
method on graphs with $N$ nodes, leading order eigensolving costs $O(N^3)$ for the
first split of a dense graph, with $O(N^2\log N)$ for the whole recursion on sparse
graphs (i.e.\ those with $m = O(N)$ edges, so that the mean degree $2m/N$ is
bounded)~\cite{newman2006modularity,newman2006finding}. Repeated bisection can also
miss the best division into more than two
groups~\cite{newman2006finding,zhang2015multiway}.
Given the above limitations, several works have developed quantum approaches to tackle
community detection, which we review in Section~\ref{ssec:q_comm_det_deteqt}.

In this work, we focus on the deteQt quantum community detection
algorithm~\cite{umeano_quantum_2025}. Unlike other quantum methods, deteQt is
a spectral modularity method, requiring fault-tolerant primitives such as the quantum
singular value transformation (QSVT)~\cite{gilyen_quantum_2019}. It prepares the
leading eigenvector of the modularity matrix $B$ as a quantum state on $\lceil \log_2 N
	\rceil$ qubits, and the signs of the amplitudes of this eigenvector label the
communities. deteQt reads the signs by amplitude flattening, using the QSVT to
approximate the sign function, and compares the resulting state against a set of probe
states. The original method used \emph{deterministic elimination}, a routine to remove
candidate community partitions exactly until a single one remained. The proposed use
case was in botnet detection, where a small community of $k$ nodes is hidden within a
much larger graph ($k \ll N/2$). However, as published it returns only a single
bipartition, and hence can split a graph into at most two communities. We describe the
algorithm in Section~\ref{ssec:q_comm_det_deteqt}.
However, our goal in this work is not directly to improve quantum community detection
algorithms, but rather to investigate their usefulness in alternative downstream data
analysis tasks, specifically in feature selection and clustering. Feature selection
removes redundant or uninformative variables before
learning~\cite{guyon2003introduction,li2017feature}. To apply community detection to
it, the features are taken as the nodes of a graph whose edges are weighted by their
pairwise correlations, so that each community is a group of correlated features.
Keeping one feature from each community then removes redundant features while every
group stays represented~\cite{kuzudisli2023review,etaiwi2025graph}. Classical detectors
such as Louvain have been used in this
way~\cite{savic2019feature,rostami2021community,feng2025corofr}. In clustering, the
nodes are data points joined in a nearest-neighbour graph and the communities are the
clusters~\cite{schaeffer2007graph}. Applying the Louvain algorithm to such graphs is
now a standard technique for grouping individual cells into cell types from their
measured gene or protein expression~\cite{levine2015data,luecken2019current}. Spectral
clustering~\cite{luxburg2007tutorial} instead partitions the data using eigenvectors of
the graph Laplacian. Spectral modularity methods are closely related: when the
eigenvector is constrained with weights given by the node degrees, the spectral
algorithm for modularity maximisation is identical to that for normalised-cut
partitioning, which spectral clustering approximates~\cite{newman2013spectral}.
Quantum algorithms have been proposed for clustering
directly~\cite{kerenidis2021spectral,kerenidis2019qmeans} and for feature selection
written in the form of a
QUBO~\cite{mucke2023feature,ferraridacrema2022towards,zoufal2023variational}. However,
to date we know of none that connects quantum community detection with these
applications.

Our contributions are to demonstrate the effectiveness of quantum community detectors
in feature selection and clustering (Section~\ref{sec:results}). We first benchmark
them on stochastic block model graphs~\cite{holland1983stochastic} with planted
communities, at signal strengths on both sides of the detectability
threshold~\cite{decelle2011asymptotic}, where quantum spectral detection can separate
from Louvain at moderate values of the ratio of intra- to inter-community connection
probabilities. For feature selection, we use the small Diabetes~\cite{smith1988using}
and Breast Cancer~\cite{street1993nuclear} datasets, and show that quantum community
detection can recover the synthetic community structure added to the Diabetes data and
select informative features, reducing random forest training time on Diabetes. We then
scale to MNIST~\cite{lecun1998gradient}, performing both feature selection over pixels,
on a $719$-node correlation graph, and direct clustering of the images, on similarity
graphs of up to $2{,}000$ nodes. Throughout, we compare against standard community
detectors: Louvain~\cite{blondel2008}, Girvan--Newman~\cite{girvan2002community},
greedy modularity~\cite{clauset2004}, Kernighan--Lin~\cite{kernighan1970efficient},
Newman's spectral method~\cite{newman2006finding} and the chi-squared,
mutual-information and ANOVA filters for feature selection.

Along the way, we build a pipeline that extends deteQt to multi-community partitions by
recursive bisection (Section~\ref{ssec:hierarchical_deteqt}), and propose projeQt, a
new quantum community detector that uses the QSVT to retain multiple eigenvectors of
the generalised modularity matrix~\cite{newman2006finding}, and can also divide a graph
into several communities from a single split (Section~\ref{ssec:deteqt_extensions}).
\section{Background}
\subsection{Community detection} \label{ssec:comm_det}
In this section, we focus on two classical methods: spectral bipartitioning of the
modularity matrix, and the Louvain algorithm, as these are the primary techniques of
relevance for this work. Throughout, we will use a particular family of models to
generate synthetic community structure: the commonly used stochastic block model
(SBM)~\cite{holland1983stochastic}. For convenience, we define it here for later
reference. An SBM on $N$-node random graphs plants $K$ synthetic communities of $N/K$
nodes each. Each node, $i$, carries a community label $\sigma_i \in \{1, \dots, K\}$,
and the adjacency matrix, $A$, is constructed probabilistically, independently for each
pair of nodes, with
\begin{equation} \label{eqn:sbm}
	\begin{aligned}
		P(A_{ij} = A_{ji} = 1) & =
		\begin{cases}
			p_{\text{in}},	& \sigma_i = \sigma_j,	  \\
			p_{\text{out}}, & \sigma_i \neq \sigma_j,
		\end{cases}
		\quad \forall\, i \neq j,			\\
		P(A_{ii} = 1)	       & = 0 \quad \forall\, i.
	\end{aligned}
\end{equation}
The ratio $r = p_{\text{in}}/p_{\text{out}}$ sets the relative strength of intra-
versus inter-community connections. For $r = 1$, each pair of nodes is connected with
the same probability, so the graph has no community structure. As $r$ grows (i.e.\ as
an edge within a community becomes more probable than one between communities), the
communities become easier to recover. The average degree of the graph is
\begin{equation} \label{eqn:sbm_degree}
	\bar c \coloneqq \mathbb{E}[k_i] = \left(\frac{N}{K} - 1\right) p_{\text{in}} +
\left(N - \frac{N}{K}\right) p_{\text{out}},
\end{equation}
where $k_i = \sum_j A_{ij}$ is the degree of node $i$.
\subsubsection{Modularity and spectral bipartitioning} \label{sssec:mod_matrix}
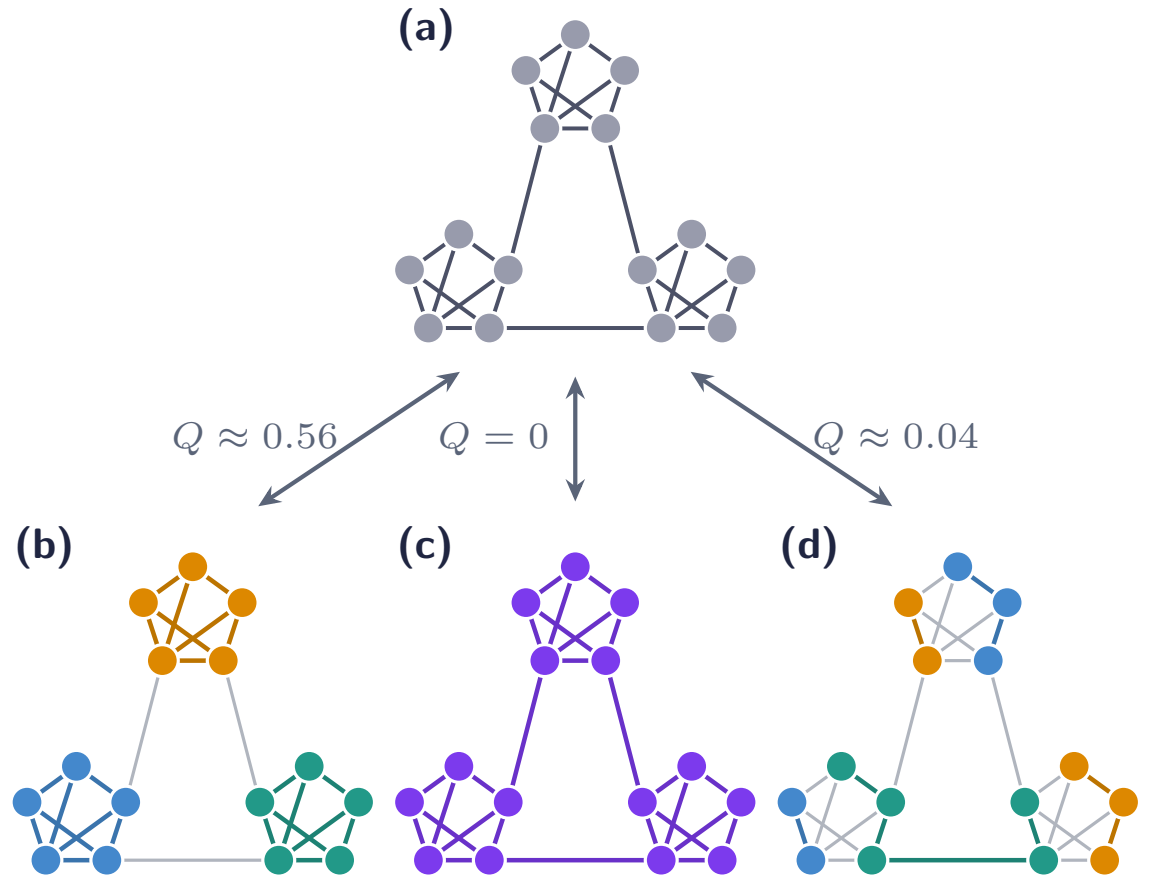
\begin{figure}[t]
	\centering

\definecolor{mtInk}{HTML}{1F2541}\definecolor{mtSlate}{HTML}{5A6478}\definecolor{mtMute}{HTML}{9CA2AE}\definecolor{mtGrey}{HTML}{6C7089}

\definecolor{c0}{HTML}{DD8800}\definecolor{c1}{HTML}{4488CC}\definecolor{c2}{HTML}{229988}\definecolor{c3}{HTML}{7C3AED}
	\begin{tikzpicture}[x=0.44cm, y=0.44cm,
		mnode/.style={circle, draw=white, line width=0.6pt, minimum size=6pt,
inner sep=0pt},
		eplain/.style={line width=0.75pt, draw=mtInk!80, line cap=round},
		ebetween/.style={line width=0.6pt, draw=mtMute!80, line cap=round},
		plab/.style={font=\sffamily\bfseries\footnotesize, text=mtInk,
anchor=north west, inner sep=0pt},
		qlab/.style={font=\sffamily\scriptsize, text=mtSlate, inner sep=1.5pt},
		parr/.style={{Stealth[length=4.2pt,
width=3.6pt]}-{Stealth[length=4.2pt, width=3.6pt]}, line width=0.8pt, draw=mtSlate,
shorten <=2pt, shorten >=2pt}]
		% Generated by a script. (a) the graph: three groups of five (5-cycle plus three chords), one bridge per pair of
		% groups. (b)-(d) the same graph under three partitions: by group; one community; each community taking two
		% nodes of one group and three of the next. An edge inside a community takes its colour; edges between are grey.
		% panel (a)
		\draw[eplain] (0.000,10.680) -- (-0.742,10.141);
		\draw[eplain] (-0.742,10.141) -- (-0.458,9.269);
		\draw[eplain] (-0.458,9.269) -- (0.458,9.269);
		\draw[eplain] (0.458,9.269) -- (0.742,10.141);
		\draw[eplain] (0.742,10.141) -- (0.000,10.680);
		\draw[eplain] (0.000,10.680) -- (-0.458,9.269);
		\draw[eplain] (-0.742,10.141) -- (0.458,9.269);
		\draw[eplain] (-0.458,9.269) -- (0.742,10.141);
		\draw[eplain] (-1.750,7.680) -- (-2.492,7.141);
		\draw[eplain] (-2.492,7.141) -- (-2.208,6.269);
		\draw[eplain] (-2.208,6.269) -- (-1.292,6.269);
		\draw[eplain] (-1.292,6.269) -- (-1.008,7.141);
		\draw[eplain] (-1.008,7.141) -- (-1.750,7.680);
		\draw[eplain] (-1.750,7.680) -- (-2.208,6.269);
		\draw[eplain] (-2.492,7.141) -- (-1.292,6.269);
		\draw[eplain] (-2.208,6.269) -- (-1.008,7.141);
		\draw[eplain] (1.750,7.680) -- (1.008,7.141);
		\draw[eplain] (1.008,7.141) -- (1.292,6.269);
		\draw[eplain] (1.292,6.269) -- (2.208,6.269);
		\draw[eplain] (2.208,6.269) -- (2.492,7.141);
		\draw[eplain] (2.492,7.141) -- (1.750,7.680);
		\draw[eplain] (1.750,7.680) -- (1.292,6.269);
		\draw[eplain] (1.008,7.141) -- (2.208,6.269);
		\draw[eplain] (1.292,6.269) -- (2.492,7.141);
		\draw[eplain] (-0.458,9.269) -- (-1.008,7.141);
		\draw[eplain] (0.458,9.269) -- (1.008,7.141);
		\draw[eplain] (-1.292,6.269) -- (1.292,6.269);
		\node[mnode, fill=mtGrey!70] at (0.000,10.680) {};
		\node[mnode, fill=mtGrey!70] at (-0.742,10.141) {};
		\node[mnode, fill=mtGrey!70] at (-0.458,9.269) {};
		\node[mnode, fill=mtGrey!70] at (0.458,9.269) {};
		\node[mnode, fill=mtGrey!70] at (0.742,10.141) {};
		\node[mnode, fill=mtGrey!70] at (-1.750,7.680) {};
		\node[mnode, fill=mtGrey!70] at (-2.492,7.141) {};
		\node[mnode, fill=mtGrey!70] at (-2.208,6.269) {};
		\node[mnode, fill=mtGrey!70] at (-1.292,6.269) {};
		\node[mnode, fill=mtGrey!70] at (-1.008,7.141) {};
		\node[mnode, fill=mtGrey!70] at (1.750,7.680) {};
		\node[mnode, fill=mtGrey!70] at (1.008,7.141) {};
		\node[mnode, fill=mtGrey!70] at (1.292,6.269) {};
		\node[mnode, fill=mtGrey!70] at (2.208,6.269) {};
		\node[mnode, fill=mtGrey!70] at (2.492,7.141) {};
		\node[plab] at (-2.7,11.10) {(a)};
		% panel (b)
		\draw[ebetween] (-6.208,1.269) -- (-6.758,-0.859);
		\draw[ebetween] (-5.292,1.269) -- (-4.742,-0.859);
		\draw[ebetween] (-7.042,-1.731) -- (-4.458,-1.731);
		\draw[line width=0.85pt, line cap=round, draw=c0!85!black]
(-5.750,2.680) -- (-6.492,2.141);
		\draw[line width=0.85pt, line cap=round, draw=c0!85!black]
(-6.492,2.141) -- (-6.208,1.269);
		\draw[line width=0.85pt, line cap=round, draw=c0!85!black]
(-6.208,1.269) -- (-5.292,1.269);
		\draw[line width=0.85pt, line cap=round, draw=c0!85!black]
(-5.292,1.269) -- (-5.008,2.141);
		\draw[line width=0.85pt, line cap=round, draw=c0!85!black]
(-5.008,2.141) -- (-5.750,2.680);
		\draw[line width=0.85pt, line cap=round, draw=c0!85!black]
(-5.750,2.680) -- (-6.208,1.269);
		\draw[line width=0.85pt, line cap=round, draw=c0!85!black]
(-6.492,2.141) -- (-5.292,1.269);
		\draw[line width=0.85pt, line cap=round, draw=c0!85!black]
(-6.208,1.269) -- (-5.008,2.141);
		\draw[line width=0.85pt, line cap=round, draw=c1!85!black]
(-7.500,-0.320) -- (-8.242,-0.859);
		\draw[line width=0.85pt, line cap=round, draw=c1!85!black]
(-8.242,-0.859) -- (-7.958,-1.731);
		\draw[line width=0.85pt, line cap=round, draw=c1!85!black]
(-7.958,-1.731) -- (-7.042,-1.731);
		\draw[line width=0.85pt, line cap=round, draw=c1!85!black]
(-7.042,-1.731) -- (-6.758,-0.859);
		\draw[line width=0.85pt, line cap=round, draw=c1!85!black]
(-6.758,-0.859) -- (-7.500,-0.320);
		\draw[line width=0.85pt, line cap=round, draw=c1!85!black]
(-7.500,-0.320) -- (-7.958,-1.731);
		\draw[line width=0.85pt, line cap=round, draw=c1!85!black]
(-8.242,-0.859) -- (-7.042,-1.731);
		\draw[line width=0.85pt, line cap=round, draw=c1!85!black]
(-7.958,-1.731) -- (-6.758,-0.859);
		\draw[line width=0.85pt, line cap=round, draw=c2!85!black]
(-4.000,-0.320) -- (-4.742,-0.859);
		\draw[line width=0.85pt, line cap=round, draw=c2!85!black]
(-4.742,-0.859) -- (-4.458,-1.731);
		\draw[line width=0.85pt, line cap=round, draw=c2!85!black]
(-4.458,-1.731) -- (-3.542,-1.731);
		\draw[line width=0.85pt, line cap=round, draw=c2!85!black]
(-3.542,-1.731) -- (-3.258,-0.859);
		\draw[line width=0.85pt, line cap=round, draw=c2!85!black]
(-3.258,-0.859) -- (-4.000,-0.320);
		\draw[line width=0.85pt, line cap=round, draw=c2!85!black]
(-4.000,-0.320) -- (-4.458,-1.731);
		\draw[line width=0.85pt, line cap=round, draw=c2!85!black]
(-4.742,-0.859) -- (-3.542,-1.731);
		\draw[line width=0.85pt, line cap=round, draw=c2!85!black]
(-4.458,-1.731) -- (-3.258,-0.859);
		\node[mnode, fill=c0] at (-5.750,2.680) {};
		\node[mnode, fill=c0] at (-6.492,2.141) {};
		\node[mnode, fill=c0] at (-6.208,1.269) {};
		\node[mnode, fill=c0] at (-5.292,1.269) {};
		\node[mnode, fill=c0] at (-5.008,2.141) {};
		\node[mnode, fill=c1] at (-7.500,-0.320) {};
		\node[mnode, fill=c1] at (-8.242,-0.859) {};
		\node[mnode, fill=c1] at (-7.958,-1.731) {};
		\node[mnode, fill=c1] at (-7.042,-1.731) {};
		\node[mnode, fill=c1] at (-6.758,-0.859) {};
		\node[mnode, fill=c2] at (-4.000,-0.320) {};
		\node[mnode, fill=c2] at (-4.742,-0.859) {};
		\node[mnode, fill=c2] at (-4.458,-1.731) {};
		\node[mnode, fill=c2] at (-3.542,-1.731) {};
		\node[mnode, fill=c2] at (-3.258,-0.859) {};
		\node[plab] at (-8.45,3.3) {(b)};
		\draw[parr] (-1.6,5.7) -- node[qlab, pos=0.5, anchor=east, xshift=-2pt]
{$Q \approx 0.56$} (-4.9,3.5);
		% panel (c)
		\draw[line width=0.85pt, line cap=round, draw=c3!85!black]
(0.000,2.680) -- (-0.742,2.141);
		\draw[line width=0.85pt, line cap=round, draw=c3!85!black]
(-0.742,2.141) -- (-0.458,1.269);
		\draw[line width=0.85pt, line cap=round, draw=c3!85!black]
(-0.458,1.269) -- (0.458,1.269);
		\draw[line width=0.85pt, line cap=round, draw=c3!85!black]
(0.458,1.269) -- (0.742,2.141);
		\draw[line width=0.85pt, line cap=round, draw=c3!85!black]
(0.742,2.141) -- (0.000,2.680);
		\draw[line width=0.85pt, line cap=round, draw=c3!85!black]
(0.000,2.680) -- (-0.458,1.269);
		\draw[line width=0.85pt, line cap=round, draw=c3!85!black]
(-0.742,2.141) -- (0.458,1.269);
		\draw[line width=0.85pt, line cap=round, draw=c3!85!black]
(-0.458,1.269) -- (0.742,2.141);
		\draw[line width=0.85pt, line cap=round, draw=c3!85!black]
(-1.750,-0.320) -- (-2.492,-0.859);
		\draw[line width=0.85pt, line cap=round, draw=c3!85!black]
(-2.492,-0.859) -- (-2.208,-1.731);
		\draw[line width=0.85pt, line cap=round, draw=c3!85!black]
(-2.208,-1.731) -- (-1.292,-1.731);
		\draw[line width=0.85pt, line cap=round, draw=c3!85!black]
(-1.292,-1.731) -- (-1.008,-0.859);
		\draw[line width=0.85pt, line cap=round, draw=c3!85!black]
(-1.008,-0.859) -- (-1.750,-0.320);
		\draw[line width=0.85pt, line cap=round, draw=c3!85!black]
(-1.750,-0.320) -- (-2.208,-1.731);
		\draw[line width=0.85pt, line cap=round, draw=c3!85!black]
(-2.492,-0.859) -- (-1.292,-1.731);
		\draw[line width=0.85pt, line cap=round, draw=c3!85!black]
(-2.208,-1.731) -- (-1.008,-0.859);
		\draw[line width=0.85pt, line cap=round, draw=c3!85!black]
(1.750,-0.320) -- (1.008,-0.859);
		\draw[line width=0.85pt, line cap=round, draw=c3!85!black]
(1.008,-0.859) -- (1.292,-1.731);
		\draw[line width=0.85pt, line cap=round, draw=c3!85!black]
(1.292,-1.731) -- (2.208,-1.731);
		\draw[line width=0.85pt, line cap=round, draw=c3!85!black]
(2.208,-1.731) -- (2.492,-0.859);
		\draw[line width=0.85pt, line cap=round, draw=c3!85!black]
(2.492,-0.859) -- (1.750,-0.320);
		\draw[line width=0.85pt, line cap=round, draw=c3!85!black]
(1.750,-0.320) -- (1.292,-1.731);
		\draw[line width=0.85pt, line cap=round, draw=c3!85!black]
(1.008,-0.859) -- (2.208,-1.731);
		\draw[line width=0.85pt, line cap=round, draw=c3!85!black]
(1.292,-1.731) -- (2.492,-0.859);
		\draw[line width=0.85pt, line cap=round, draw=c3!85!black]
(-0.458,1.269) -- (-1.008,-0.859);
		\draw[line width=0.85pt, line cap=round, draw=c3!85!black]
(0.458,1.269) -- (1.008,-0.859);
		\draw[line width=0.85pt, line cap=round, draw=c3!85!black]
(-1.292,-1.731) -- (1.292,-1.731);
		\node[mnode, fill=c3] at (0.000,2.680) {};
		\node[mnode, fill=c3] at (-0.742,2.141) {};
		\node[mnode, fill=c3] at (-0.458,1.269) {};
		\node[mnode, fill=c3] at (0.458,1.269) {};
		\node[mnode, fill=c3] at (0.742,2.141) {};
		\node[mnode, fill=c3] at (-1.750,-0.320) {};
		\node[mnode, fill=c3] at (-2.492,-0.859) {};
		\node[mnode, fill=c3] at (-2.208,-1.731) {};
		\node[mnode, fill=c3] at (-1.292,-1.731) {};
		\node[mnode, fill=c3] at (-1.008,-0.859) {};
		\node[mnode, fill=c3] at (1.750,-0.320) {};
		\node[mnode, fill=c3] at (1.008,-0.859) {};
		\node[mnode, fill=c3] at (1.292,-1.731) {};
		\node[mnode, fill=c3] at (2.208,-1.731) {};
		\node[mnode, fill=c3] at (2.492,-0.859) {};
		\node[plab] at (-2.70,3.3) {(c)};
		\draw[parr] (0,5.7) -- node[qlab, pos=0.5, anchor=east, xshift=-3pt]
{$Q = 0$} (0,3.5);
		% panel (d)
		\draw[ebetween] (5.750,2.680) -- (5.008,2.141);
		\draw[ebetween] (5.292,1.269) -- (6.208,1.269);
		\draw[ebetween] (5.750,2.680) -- (5.292,1.269);
		\draw[ebetween] (5.008,2.141) -- (6.208,1.269);
		\draw[ebetween] (5.292,1.269) -- (6.492,2.141);
		\draw[ebetween] (4.000,-0.320) -- (3.258,-0.859);
		\draw[ebetween] (3.542,-1.731) -- (4.458,-1.731);
		\draw[ebetween] (4.000,-0.320) -- (3.542,-1.731);
		\draw[ebetween] (3.258,-0.859) -- (4.458,-1.731);
		\draw[ebetween] (3.542,-1.731) -- (4.742,-0.859);
		\draw[ebetween] (7.500,-0.320) -- (6.758,-0.859);
		\draw[ebetween] (7.042,-1.731) -- (7.958,-1.731);
		\draw[ebetween] (7.500,-0.320) -- (7.042,-1.731);
		\draw[ebetween] (6.758,-0.859) -- (7.958,-1.731);
		\draw[ebetween] (7.042,-1.731) -- (8.242,-0.859);
		\draw[ebetween] (5.292,1.269) -- (4.742,-0.859);
		\draw[ebetween] (6.208,1.269) -- (6.758,-0.859);
		\draw[line width=0.85pt, line cap=round, draw=c0!85!black]
(5.008,2.141) -- (5.292,1.269);
		\draw[line width=0.85pt, line cap=round, draw=c1!85!black]
(6.208,1.269) -- (6.492,2.141);
		\draw[line width=0.85pt, line cap=round, draw=c1!85!black]
(6.492,2.141) -- (5.750,2.680);
		\draw[line width=0.85pt, line cap=round, draw=c1!85!black]
(3.258,-0.859) -- (3.542,-1.731);
		\draw[line width=0.85pt, line cap=round, draw=c2!85!black]
(4.458,-1.731) -- (4.742,-0.859);
		\draw[line width=0.85pt, line cap=round, draw=c2!85!black]
(4.742,-0.859) -- (4.000,-0.320);
		\draw[line width=0.85pt, line cap=round, draw=c2!85!black]
(6.758,-0.859) -- (7.042,-1.731);
		\draw[line width=0.85pt, line cap=round, draw=c0!85!black]
(7.958,-1.731) -- (8.242,-0.859);
		\draw[line width=0.85pt, line cap=round, draw=c0!85!black]
(8.242,-0.859) -- (7.500,-0.320);
		\draw[line width=0.85pt, line cap=round, draw=c2!85!black]
(4.458,-1.731) -- (7.042,-1.731);
		\node[mnode, fill=c1] at (5.750,2.680) {};
		\node[mnode, fill=c0] at (5.008,2.141) {};
		\node[mnode, fill=c0] at (5.292,1.269) {};
		\node[mnode, fill=c1] at (6.208,1.269) {};
		\node[mnode, fill=c1] at (6.492,2.141) {};
		\node[mnode, fill=c2] at (4.000,-0.320) {};
		\node[mnode, fill=c1] at (3.258,-0.859) {};
		\node[mnode, fill=c1] at (3.542,-1.731) {};
		\node[mnode, fill=c2] at (4.458,-1.731) {};
		\node[mnode, fill=c2] at (4.742,-0.859) {};
		\node[mnode, fill=c0] at (7.500,-0.320) {};
		\node[mnode, fill=c2] at (6.758,-0.859) {};
		\node[mnode, fill=c2] at (7.042,-1.731) {};
		\node[mnode, fill=c0] at (7.958,-1.731) {};
		\node[mnode, fill=c0] at (8.242,-0.859) {};
		\node[plab] at (3.05,3.3) {(d)};
		\draw[parr] (1.6,5.7) -- node[qlab, pos=0.5, anchor=west, xshift=2pt]
{$Q \approx 0.04$} (4.9,3.5);
	\end{tikzpicture}
	\caption{Modularity scores a partition of a graph by the number of edges that
fall inside its communities, compared with the number that would be expected by random
association, i.e.\ in a graph with the same node degrees but randomly permuted
connections. \textbf{(a)}~The graph, with $27$ edges. \textbf{(b)--(d)}~Three
partitions of it, each community in its own colour: an edge between two members of the
same community takes their colour, and an edge between communities is grey.
\textbf{(b)}~High modularity: the partition follows the three densely linked groups,
and $24$ edges lie inside a community, against $9$ expected by random permutation.
\textbf{(c)}~A single community containing all $27$ edges, the same number expected at
random. \textbf{(d)}~The partition is spread across the densely connected physical
groups: only $10$ edges lie inside a community, against $9$ by chance.}
	\label{fig:modularity_toy}
\end{figure}
Modularity, introduced in Ref.~\cite{newman2004finding}, measures how much the
fraction of edges inside the communities of a proposed partition exceeds the fraction
expected under a degree-preserving null model, in which edges are placed at random with
every node's degree kept fixed. For a graph \(\mathcal{G}\) with \(N\) nodes, \(m\)
edges, adjacency matrix $A_{ij}$, and node degrees $k_i$, the modularity of a partition
is
\begin{equation} \label{eqn:modularity_fun}
	Q = \frac{1}{2m} \sum_{i,j} \left( A_{ij} - \frac{k_i k_j}{2m} \right)
	\delta(c_i, c_j),
\end{equation}
where $\delta(c_i,c_j) = 1$ if nodes $i$ and $j$ are assigned to the same community and
0 otherwise. Figure~\ref{fig:modularity_toy} illustrates modularity on a small graph.
The quantity $k_i k_j / 2m$ is the expected number of edges between $i$ and $j$ in the
null model, so the entries of the \emph{modularity matrix},
\begin{equation}\label{eqn:modularity_mat}
	B_{ij} = A_{ij} - \frac{k_i k_j}{2m},
\end{equation}
are positive when a pair of nodes is connected more densely than expected and negative
otherwise. For a bipartition encoded as $\boldsymbol{s} \in \{\pm 1\}^N$, modularity
takes the quadratic form~\cite{newman2006modularity}
\begin{equation} \label{eqn:modularity_spectral}
	Q = \frac{1}{4m} \boldsymbol{s}^{\top} B \boldsymbol{s}.
\end{equation}
Maximising this over $\{\pm 1\}^N$ is NP-hard~\cite{brandes2008modularity}. The
standard relaxation replaces the integrality constraint with $\|\boldsymbol{s}\| = 1$.
The maximum of $\boldsymbol{s}^\top B \boldsymbol{s}$ subject to this constraint is
attained by the leading eigenvector of $B$ with value $\lambda_{\max}(B)$. Rounding its
components by sign recovers an approximate bipartition.
\subsubsection{Generalised modularity matrix} \label{sssec:gen_mod_matrix}
When attempting to use modularity to perform recursive splitting of a previously found
community, $g$, we encounter an issue. Specifically, if we were to treat $g$ as a new
graph and optimise its modularity, we would not optimise the modularity of the global
graph, because we would not account for connections between $g$ and nodes outside
$g$~\cite{newman2006modularity,newman2006finding}. Newman's generalised modularity
matrix~\cite{newman2006finding} instead scores a division of $g$ by the change it makes
to the modularity of the whole graph:
\begin{equation} \label{eqn:gen_modularity_mat}
	B^{(g)}_{ij} = A_{ij} - \frac{k_i k_j}{2m} - \delta_{ij} \left( k_i^{(g)} -
\frac{k_i \, d_g}{2m} \right),
\end{equation}
where $k_i^{(g)} = \sum_{j \in g} A_{ij}$ is the number of edges from node $i$ into $g$
and $d_g = \sum_{j \in g} k_j$ is the total degree of $g$. The diagonal term makes
every row sum to zero over $g$,
\begin{equation} \label{eqn:gen_mod_rowsum}
	\sum_{j \in g} B^{(g)}_{ij} = 0.
\end{equation}
so that dividing $g$ in two with signs $\boldsymbol{s} \in \{\pm 1\}^{N_g}$ changes the
modularity of the whole graph by $\boldsymbol{s}^{\top} B^{(g)} \boldsymbol{s}/4m$, and
the leading eigenvector of $B^{(g)}$ gives the best division under the same relaxation
as for $B$~\cite{newman2006finding}. When $g$ is the whole graph, $B^{(g)}$ reduces to
$B$. A division is kept only if it raises the modularity, for which a positive leading
eigenvalue is necessary but not sufficient~\cite{newman2006modularity}.
\subsubsection{Limitations of spectral modularity methods}
\label{sssec:spectral_limits}
As mentioned above, spectral methods have limitations which are of relevance to this
work. The first is the \emph{resolution limit}~\cite{fortunato2007resolution}. By
Eq.~\eqref{eqn:modularity_fun}, merging two communities $A$ and $B$ changes the
modularity by
\begin{equation} \label{eqn:merge_dq}
	\Delta Q = \frac{l_{AB}}{m} - \frac{d_A d_B}{2m^2},
\end{equation}
where $l_{AB}$ is the number of edges between them and $d_A$, $d_B$ are their total
degrees. If we have two communities, each with $l$ internal edges and a single edge
between them (such that $d_A \approx d_B \approx 2l$), a merge of $A$ and $B$ will
raise the modularity whenever $l < \sqrt{m/2}$. As a result, small communities end up
merging regardless of how densely connected they are internally.
The second is that relating community structure to the leading eigenvector of a
modularity matrix is only meaningful when a certain amount of structure exists. For
spectral methods to be useful, the leading eigenvalue must be larger than that of a
random graph with the same degree structure but no communities. To illustrate this,
take a graph with adjacency matrix $A$ drawn from the SBM of Eq.~\eqref{eqn:sbm} with
$K = 2$, $p_{\text{in}} = c_{\mathrm{in}}/N$ and $p_{\text{out}} = c_{\mathrm{out}}/N$,
so that $\bar c \simeq (c_{\mathrm{in}} + c_{\mathrm{out}})/2$, and let $B$ be its
modularity matrix, Eq.~\eqref{eqn:modularity_mat}. The expected modularity matrix,
$\mathbb{E}[B]$, has a single nonzero eigenvalue, $\bar\lambda = (c_{\mathrm{in}} -
	c_{\mathrm{out}})/2$, whose eigenvector signs,
$\mathrm{sign}(\boldsymbol{v}_{\bar\lambda})$, label the communities. Write $B =
	\mathbb{E}[B] + X$, the expected matrix plus a zero-mean random matrix $X
\coloneqq B -
	\mathbb{E}[B]$. By Eq.~\eqref{eqn:modularity_mat},
\begin{equation} \label{eqn:sbm_noise}
	X_{ij} = \underbrace{A_{ij} - \mathbb{E}[A_{ij}]}_{\text{independent entries}}
	- \underbrace{\left( \frac{k_i k_j}{2m} - \mathbb{E}\!\left[ \frac{k_i k_j}{2m}
\right] \right)}_{\text{low rank}},
\end{equation}
where the second term does not change the range of eigenvalues of
$X$~\cite{nadakuditi2012graph}. The entries of $A - \mathbb{E}[A]$ above the diagonal
are independent with mean zero, and each of its rows has total variance $\sum_j
\mathrm{Var}(A_{ij}) \simeq \bar c$. When $\bar c \gg \log N$, such a matrix has
spectral norm $\lVert A - \mathbb{E}[A] \rVert \simeq 2\sqrt{\bar
c}$~\cite{bandeira2016sharp}, so all the eigenvalues of $X$ lie in $[-2\sqrt{\bar c},
2\sqrt{\bar c}]$. A random graph with the same mean degree has $\mathbb{E}[B] \approx
0$, and therefore $\lambda_{\max}(B) \simeq 2\sqrt{\bar c}$. With communities, the
leading eigenvalue is~\cite{nadakuditi2012graph}
\begin{equation} \label{eqn:sbm_outlier}
	\lambda_{\max}(B) \simeq
	\begin{cases}
		\bar\lambda + \bar c/\bar\lambda, & \bar\lambda > \sqrt{\bar c},   \\
		2\sqrt{\bar c}, 		  & \bar\lambda \le \sqrt{\bar c},
	\end{cases}
\end{equation}
which exceeds that of the random graph only when $\bar\lambda > \sqrt{\bar c}$, that is
when $c_{\mathrm{in}} - c_{\mathrm{out}} > 2\sqrt{\bar c}$. The point $c_{\mathrm{in}}
- c_{\mathrm{out}} = 2\sqrt{\bar c}$ is the detectability
threshold~\cite{decelle2011asymptotic}. If $c_{\mathrm{in}} - c_{\mathrm{out}} <
2\sqrt{\bar c}$, no spectral approach based on the modularity matrix can find the
communities. Later in this work, we will use multiple leading eigenvectors to extract
community structure. In this example, the eigenvectors with eigenvalues above
$2\sqrt{\bar c}$ are correlated with the communities, and each of those below it
carries no information about them.
\subsubsection{The Louvain algorithm} \label{sssec:louvain}
Spectral bipartitioning requires the leading eigenvector of $B^{(g)}$ for each group
$g$ it splits, at every level of the recursion. The Louvain
algorithm~\cite{blondel2008} is faster. It optimises modularity through two alternating
greedy phases.
In the first phase, each node $i$ is considered in turn and tentatively moved to each
neighbouring community $C$. The modularity gain from moving node $i$ into community $C$
is
\begin{equation} \label{eqn:louvain_delta_q}
	\Delta Q_i = \frac{1}{m}\left[ k_{i,\mathrm{in}}^{(C)} - \frac{k_i \,
\Sigma_C}{2m} \right],
\end{equation}
where $k_{i,\mathrm{in}}^{(C)} = \sum_{j \in C} A_{ij}$ is the sum of edge weights from
$i$ to nodes in $C$, $k_i$ is the degree of node $i$, and $\Sigma_C = \sum_{j \in C}
	k_j$ is the total degree of community $C$. Removing $i$ from its current
community
$C_0$ changes the modularity by $-\Delta Q_i$ evaluated for $C_0 \setminus \{i\}$, so
the net gain of the move is $\Delta Q_i(C) - \Delta Q_i(C_0 \setminus \{i\})$. Node $i$
is reassigned to the community $C^* = \arg\max_C \Delta Q_i$ that yields the largest
positive net gain, or left in place if no move increases $Q$. This sweep is repeated
until no single-node move improves modularity. In the second phase, each community is
contracted into a single super-node: self-loops represent the internal edge weight and
inter-community edges are summed. The first phase then repeats on this coarsened graph.
The two phases alternate until the partition stabilises. Its running time appears
linear in the number of edges on sparse graphs~\cite{blondel2008,traag2015faster}, and
the algorithm handles networks with millions of nodes~\cite{blondel2008}. The Leiden
algorithm~\cite{traag2019louvain} refines Louvain by guaranteeing that every community
in the final partition is connected, a property Louvain does not enforce.
\subsubsection{Limitations of the Louvain algorithm} \label{sssec:louvain_limits}
Since the Louvain algorithm also maximises modularity, it inherits the resolution limit
discussed above. It also has two further limitations. The first is that, due to its
greedy nature, the partition selected depends strongly on the order in which nodes are
visited~\cite{blondel2008}. A node move in the first phase is accepted if it raises
$Q$, as discussed above. As such, different node orderings can achieve different local
modularity maxima, and modularity typically has many such local maxima with nearly
equal values~\cite{good2010performance}. Thus, different runs can return different
partitions with similar modularity. In contrast, spectral bisection depends on the
graph alone. When $\lambda_{\max}(B)$ is non-degenerate, the partition is unique.
The second is that, even on graphs drawn from Eq.~\eqref{eqn:sbm}, no condition has
been proven under which the standard Louvain algorithm's partition reflects the
communities; the one known guarantee is for a modified two-group
version~\cite{cohenaddad2020power}. Even a random graph with no communities,
Eq.~\eqref{eqn:sbm} with $r = 1$, has partitions of high
modularity~\cite{guimera2004modularity}, and Louvain returns one, even if it is
meaningless. Spectral bisection also returns a partition here, but for two groups
Eq.~\eqref{eqn:sbm_outlier} gives the condition under which it is correlated with the
communities, $c_{\mathrm{in}} - c_{\mathrm{out}} > 2\sqrt{\bar c}$.
Section~\ref{ssec:results_sbm} compares the two on graphs drawn from
Eq.~\eqref{eqn:sbm}.
\subsection{Quantum community detection and deteQt} \label{ssec:q_comm_det_deteqt}
Most quantum approaches to community detection to date have followed one of two
approaches: either encoding the modularity matrix in a diagonal Hamiltonian and finding
its ground state via quantum
annealing~\cite{ushijima2017graph,negre2020detecting,zahedinejad2019multi,fernandezcampoamor2021community,falcoroget2026modularity,bucher2025evaluating}
or a variational circuit~\cite{shaydulin2019network,gemeinhardt2021}, or Louvain-based
algorithms with fault-tolerant routines based on Grover search~\cite{cade2024}. Besides
these, some methods use quantum walks, assigning nodes to communities from the walk's
probability distribution~\cite{mukai2020discrete,rahaman2026community}, and
Ref.~\cite{schmidhuber2025} showed that a quartic speedup is possible when deciding
whether a hypergraph has community structure, by estimating the largest eigenvalue of a
matrix built from the hypergraph, the Kikuchi matrix, with quantum phase estimation.
In this work, we focus on the deteQt quantum algorithm, which is a community recovery
(not just existence detection) algorithm based on spectral features of the modularity
matrix. Instead of an Ising-like encoding on $N$ qubits, deteQt uses an encoding of the
leading eigenvector on $\lceil\log_2 N\rceil$ qubits as the solution of the relaxed
problem of Eq.~\eqref{eqn:modularity_spectral}. The algorithm contains the following
ingredients and steps.
\paragraph{Signed leading eigenvector}
The first step is to prepare the leading eigenvector of $B$, $\ket{\lambda_{\max}}$. To
access it on a quantum computer, the problem is recast as ground-state preparation,
finding the state that minimises $\langle \psi | (-B) | \psi \rangle$, which is done
via a ground-state preparation routine. Ref.~\cite{umeano_quantum_2025} takes
VQE~\cite{peruzzo_variational_2014} as an example routine, which produces a
parameterised unitary $U_{\max}(\boldsymbol{\theta})$ on $n = \lceil\log_2 N\rceil$
qubits with $U_{\max}(\boldsymbol{\theta}^*)\ket{0}^{\otimes n} \approx
	\ket{\lambda_{\max}}$.
The next step converts the real-valued amplitudes of this state into $\{\pm 1\}$ signs
that label the two communities. Given controlled access to
$U_{\max}(\boldsymbol{\theta}^*)$, the methods of Ref.~\cite{rattew_non-linear_2024}
construct a block encoding of the diagonal matrix $A = \text{diag}(a_1, \dots,
	a_{2^n})$ of amplitudes. The construction uses $O(1)$ calls to controlled
$U_{\max}$. A
block encoding of a matrix $M$ is a unitary whose top-left block is
$M/\alpha$~\cite{low_hamiltonian_2019,gilyen_quantum_2019}. Applying the
QSVT~\cite{gilyen_quantum_2019} with a sign-function polynomial of degree $d$ yields
$\tilde{A} = \text{diag}(\text{sign}(a_1), \dots, \text{sign}(a_{2^n}))$. Applied to
the uniform superposition $H^{\otimes n}\ket{0}$ of the $n$ system qubits, the QSVT
circuit's branch with its ancillas in $\ket{0}$ is a real equally weighted (REW) state,
in which every amplitude is flattened to its sign
\begin{equation} \label{eq:signed_eigenvector}
	\ket{\boldsymbol{s}} \approx \frac{1}{\sqrt{2^n}}\sum_{i=1}^{2^n} s_i \ket{i},
\end{equation}
where $s_i = \text{sign}(a_i) \in \{+1, -1, 0\}$ gives the community label of node $i$
for $i \leq N$, and $s_i = 0$ for padded indices $i > N$ when $N$ is not a power of
two. The polynomial degree $d$ controls the sharpness of the sign approximation near
zero. Larger $d$ improves fidelity at the cost of deeper
circuits~\cite{low_hamiltonian_2019}.
\paragraph{Probe states and the LCU}
Once the signed eigenvector is prepared, the second stage identifies which bipartition
it encodes. The candidate set $\mathcal{G}_{\text{cand}}$ consists of all
$\binom{N}{k}$ sign vectors with exactly $k$ entries equal to $+1$, representing every
possible assignment of $k$ nodes to one community. To distinguish among these, a
complementary set of \emph{probe} states $\mathcal{G}_{\text{LCU}}$ is constructed from
hypergraph states~\cite{rossi_quantum_2013} at Hamming weight $k' = N/2 - k$. The key
property is that each incorrect candidate has zero overlap with at least one probe. The
number of probes is $|\mathcal{G}_{\text{LCU}}| = \binom{N}{k'}$, and they are embedded
into a linear combination of unitaries (LCU):
\begin{equation} \label{eq:lcu_probes}
	\ket{\psi_{\text{hyp}}} = \frac{1}{\sqrt{|\mathcal{G}_{\text{LCU}}|}}\sum_{x
\in \mathcal{G}_{\text{LCU}}} \ket{x}\ket{G(x)},
\end{equation}
where $\ket{G(x)}$ is the hypergraph state for probe $x$. Each measurement of the
ancilla returns one probe index $x$, with probability proportional to
$\lvert\braket{G(x)}{\boldsymbol{s}}\rvert^2$, the squared overlap of that probe with
the signed eigenvector.
\paragraph{Deterministic elimination}
deteQt uses this zero overlap feature for a deterministic elimination strategy as
follows. Given a probe index $x$, every candidate $G_c \in \mathcal{G}_{\text{cand}}$
satisfying $|\braket{G_c}{G(x)}|^2 = 0$ (or below a threshold) is eliminated, since it
could not have produced that outcome. Repeating this with fresh probe samples narrows
the candidate set until one assignment remains or the budget is exhausted. This
succeeds because the correct partition has non-zero overlap with every probe that can
be sampled, while each incorrect candidate has zero overlap with at least one. The
number of rounds therefore scales as $\log_2 \lvert\mathcal{G}_{\text{cand}}\rvert =
	\log_2 \binom{N}{k}$, on the assumption in Ref.~\cite{umeano_quantum_2025} that
each
sample eliminates about half of the remaining candidates. Loosening the zero-overlap
requirement leads to the second case considered by Ref.~\cite{umeano_quantum_2025},
which is the case of \emph{small-overlap}. It takes $\binom{N}{k_{\mathrm{LCU}}}$
probes of a fixed (small) Hamming weight, $k_{\mathrm{LCU}}$, so for $k_{\mathrm{LCU}}
	= 1$ we have $\binom{N}{1} = N$ probes. It then uses these probes to eliminate
candidates whose overlap with a sampled probe falls below a certain threshold.
\subsection{Feature selection} \label{ssec:feature_selection}
The first of the two data-analysis tasks to which this work applies community detection
is feature selection. Feature selection chooses a subset of the \(N\) available input
variables that retains predictive or analytical value. When \(N\) is large relative to
the number of samples, fitting all features risks overfitting and adds unnecessary
computational cost~\cite{guyon2003introduction,li2017feature}. The standard taxonomy
distinguishes 1) filter methods, which score features by statistical criteria (mutual
information~\cite{battiti1994using,peng2005feature}, correlation, chi-squared tests)
without reference to a learning algorithm, but miss multivariate interactions; 2)
wrapper methods, which evaluate candidate subsets by cross-validated model
performance~\cite{kohavi1997wrappers,guyon2003introduction}, but face a \(2^N\) search
space; and 3) embedded methods, which perform selection during training itself
(\(\ell_1\)-regularised
regression~\cite{tibshirani1996regression,zou2005regularization}, tree-based
importance~\cite{breiman2001random}), but couple the selected features to the inductive
bias of a single model class~\cite{li2017feature}. With the exception of
redundancy-aware filters such as mRMR~\cite{peng2005feature}, the conditional criteria
JMI and CMIM~\cite{brown2012conditional} and FCBF~\cite{yu2004efficient}, which
penalise dependence between features within the scoring function, these families do not
represent the correlation structure among features explicitly. Graph-based methods,
described next, do represent this structure, and they are the basis of the methods in
this work.
\subsubsection{Community detection for feature selection} \label{ssec:cd_fs}
An alternative formulation for feature selection constructs a graph whose nodes are
features and whose edges are weighted by pairwise similarity (correlation, mutual
information, fuzzy-rough dependency, or
similar)~\cite{kuzudisli2023review,etaiwi2025graph}. Related spectral methods such as
the Laplacian score~\cite{he2005laplacian} and the SPEC
framework~\cite{zhao2007spectral} rank features using a sample-similarity graph rather
than a feature graph, but establish the broader connection between graph spectra and
feature relevance. Grouping methods partition the feature graph and keep one or more
representatives from each group.
This graph-based feature selection mentality can be directly incorporated with
community detection. For example, Ref.~\cite{savic2019feature} builds a weighted
Spearman correlation network over the features whose mutual information with the class
passes a threshold, adds edges until the graph is connected, detects communities, and
keeps one or more features from each community. Ref.~\cite{rostami2021community}
weights the edges by Pearson correlation, finds the communities with the ISCD
algorithm, and uses them to constrain a genetic algorithm that takes a specified number
of features from every community. Ref.~\cite{feng2025corofr} builds a feature
interaction graph from a robust fuzzy-rough model, runs the Louvain
algorithm~\cite{blondel2008} on it, and ranks the features within each community with
the same model. In bioinformatics, weighted gene co-expression network analysis
(WGCNA)~\cite{langfelder2008wgcna} applies a similar pipeline to gene correlation
networks and has become a standard preprocessing step for high-dimensional omics data.
It soft-thresholds the correlations~\cite{zhang2005general}, finds modules by
hierarchical clustering~\cite{langfelder2008dynamic}, and summarises each module by its
eigengene or by an intramodular hub gene~\cite{langfelder2013hub}. With adjacency
$|r_{ij}|$, the hub gene is the feature whose mean absolute correlation with the other
features of its module is largest. Equivalently, the hub gene has the minimum average
distance (MAD) to the other features of its module, for the distance $1 - |r_{ij}|$. We
use this feature as a representative in this work. Ref.~\cite{macmahon2015community}
shows that community detection on correlation matrices with the usual null model is
biased, and replaces the null with one from random-matrix theory. This applies to our
pipeline, which uses the usual null on a thresholded correlation graph
(Section~\ref{sssec:graph_construction}).
The success of these methods depends on the quality of the underlying community
detection, and as such any improvement to the community detection algorithms can also
improve many downstream tasks and applications.
\section{Pipeline for data analysis} \label{sec:deteqt_data_analysis}

We define one shared pipeline for multiple possible data analysis tasks. A graph is built from the
data (Section~\ref{sssec:graph_construction}). deteQt
(Section~\ref{ssec:hierarchical_deteqt}) or projeQt
(Section~\ref{ssec:deteqt_extensions}) quantum community detectors divides it into $K$ communities, possibly with 
classical refinement. A representative feature from each community, or the clusters themselves, are returned
for feature selection or clustering, respectively (Section~\ref{ssec:pipeline_output}).
Both methods maximise the modularity of Eq.~\eqref{eqn:modularity_fun} and differ in
how they read a partition out of the spectrum of $B$. deteQt reads a bisection from
probe states on one eigenvector and recurses. projeQt keeps the part of the spectrum
above a cut and divides a group by kernel $k$-means on the filtered states
(Figure~\ref{fig:hero}, Appendix~\ref{app:pipeline}).
\subsection{Graphs} \label{sssec:graph_construction}
\paragraph{Feature graph} For feature selection the nodes are the features, and an edge
connects two features
when their Pearson correlation exceeds a threshold. A fixed threshold has two problems.
First, it does not adapt to the sample size: a correlation of \(r = 0.1\) is noise in a
dataset of \(50\) samples but highly significant in one of \(10{,}000\). Second, a
user-defined threshold is a hyperparameter that can change downstream performance if
poorly chosen.
We address this by deriving the threshold from the two-tailed \(t\)-test for Pearson
correlation. Given a sample of size \(n\), the test statistic for a correlation
coefficient \(r\) is
\begin{equation} \label{eq:corr_t_stat}
	t = r \sqrt{\frac{n - 2}{1 - r^2}},
\end{equation}
with \(n - 2\) degrees of freedom. Inverting this at significance level \(\alpha\)
yields the minimum correlation that is statistically distinguishable from zero:
\begin{equation} \label{eq:critical_r}
	r_{\mathrm{crit}} = \frac{t_{\mathrm{crit}}(\alpha,\, n-2)}
	{\sqrt{t_{\mathrm{crit}}^2(\alpha,\, n-2) + (n-2)}},
\end{equation}
where \(t_{\mathrm{crit}}(\alpha, n-2)\) is the two-tailed critical value of the
\(t\)-distribution. We set \(\alpha = 0.05\) throughout. An edge is included in the
feature graph \(\mathcal{G}\) between features \(i\) and \(j\) whenever \(|r_{ij}| \geq
r_{\mathrm{crit}}\), with weight \(|r_{ij}|\). Note that \(|r_{ij}| = 1\) involves a perfectly correlated pairwise comparison between feature $i$ and $j$. This produces denser graphs for larger
datasets, where weak correlations carry genuine signal, and sparser graphs for small
datasets, where they do not. The correlation matrix is computed using all available
samples, without using label information. Ref.~\cite{macmahon2015community} shows that
the usual null model of modularity is biased on correlation matrices. We keep the usual
null. The threshold removes only the edges that cannot be distinguished from noise
(Section~\ref{ssec:results_diabetes}).
On datasets where weak correlations mask community structure, a split may be accepted
but be poor, because noisy edges distort the eigenvector. The pipeline therefore runs on
	the feature graph with edges below a weight threshold $\tau$ removed, starting
from
	$\tau_0$, and treats the remaining edges as unweighted. Its acceptance test
	(Section~\ref{ssec:hierarchical_deteqt}) can reject a split.
When a rejection happens, $\tau$ is increased by $0.1$ and the subgraph is re-tested
with weak edges removed. If $\tau$ exceeds a ceiling ($\tau > 1.0$), the community is
kept unsplit and $\tau$ returns to $\tau_0$. It is not reset after a successful split,
so a raised threshold carries on to the next group. This retry loop can expose
structure that is invisible in the dense graph. We treat $\tau_0$ as a hyperparameter
whose default is $r_{\mathrm{crit}}$ (Eq.~\eqref{eq:critical_r}).
Appendix~\ref{app:thres_study} compares the influence of two  values of $\tau_0$, $r_{\mathrm{crit}}$ and $0.95$, for a small dataset.
% Section~\ref{sssec:graph_construction} Figure~\ref{fig_threshold_comp} Section~\ref{ssec:results_sbm}
%
\paragraph{Data-point graph} For clustering problems, the nodes are the data points
(generalisable for any kind of
data), each connected to its nearest neighbours, with unweighted edges (weight 1)
(Section~\ref{ssec:results_mnist_clustering}). Correctly defining the neighbourhood of
each node will alter the results. Since every edge has weight $1$, $\tau_0 = 0$ and the
threshold has no effect there (Section~\ref{ssec:results_sbm}).
\subsection{deteQt for \texorpdfstring{$K$}{K} communities} \label{ssec:hierarchical_deteqt}
\paragraph{One bisection}
 deteQt divides a group in two by the protocol of
Section~\ref{ssec:q_comm_det_deteqt},
with the $N$ single-node probes of the small-overlap case, $k_{\mathrm{LCU}} = 1$. The
leading eigenvector of the group's modularity matrix is prepared by a VQE circuit
$U_{\max}$. For all small scale results in this work, we directly simulate the circuits,
resorting to classical emulation of the effective action of each circuits when the problem size
grows. The amplitudes of the prepared state $a_i$ are block encoded into a diagonal matrix as in~\cite{umeano_quantum_2025} and then a QSVT is used to apply an odd polynomial of degree $d = 5$ fitted to $\tanh(10x)$ to approximate the sign function, $\texttt{sign}(a_i)\in \{\pm 1\}$. We restrict to $k_{\mathrm{LCU}}=1$ in this work, so each probe corresponds to a particular Hamming weight $=1$ state. As such, probe $j$, the
hypergraph state with node $j$ alone flipped, is drawn with probability $p_j \propto
\lvert\braket{G_j}{\psi}\rvert^2$. We take $S = 1000$ samples.
For each node $i$, the frequency $\nu_i$ is defined as
\begin{equation} \label{eq:node_frequency}
\nu_i := \sum_{s=1}^{S} \mathds{1}[j_s=i],
\end{equation}
% %
Each measurement returns a probe index $j_s$, not a community label. Because
probe $i$ corresponds to a HW$=1$ state with qubit (node) $i$ flipped, $\nu_i$
counts how often the $S$ draws return probe $i$, and $\mathbb{E}[\nu_i]=S p_i$.
The `correct' community size $k^*$ (which is distinct from the probe size, $k_{\mathrm{LCU}}$) can be estimated as described in
Ref.~\cite{umeano_quantum_2025}. deteQt ranks nodes by $\nu_i$, assigns the
$k^*$ highest-frequency nodes to one community, and assigns the
remaining nodes to the other.
\paragraph{Acceptance}
Since the measurement procedure is not deterministic, we attach a gating mechansim to detect the statistical significance of a proposed community. Once a community partition is proposed, deteQt tests whether its probe
distribution differs from uniform. For probe $j$, the expected count among $S$
draws is $S p_j$, while the uniform expected count is $S/N$. We use the $\chi^2$
goodness-of-fit test to compare the estimated distribution with uniform, and the partition is accepted when its
$p$-value is at most $0.05$. A distribution sufficiently close to uniform is not a reliable predictor of communities so the split is rejected. In this case, we increase the threshold, $\tau$ by $\Delta \tau$ to remove weaker edges then repeat the process. If $\tau$ reaches $1.0$, the graph is kept unsplit. A community consisting of all graph nodes (i.e. $k^*=0$) is also refused.

In all experiments in this work, we use $S=1000$ samples. With $k_{\mathrm{LCU}}=1$ and
exact signs, probe $j$ has sign $+1$ at every node except $j$, giving overlap
$\langle \pi_j | \boldsymbol{s}\rangle = \sigma - 2s_j$, where
$\sigma=\sum_i s_i$. For a balanced split, $\sigma=0$, so all probes have the
same overlap magnitude and the distribution is uniform. Exact sign rounding
therefore causes the test to reject a balanced split. The degree-$5$ sign
approximation retains amplitude magnitudes and breaks this degeneracy.
\paragraph{Recursion}
To obtain $K$ communities, deteQt recursively applies the accepted bisection. It
keeps candidate groups in a first-in, first-out queue, splitting the next group and
adding its two parts to the back. A group that cannot be split after the threshold
retry is finalised. Recursion ends when the queue and finalised groups together
contain $K$ communities, or when the queue is empty.

At each split of group $g$, deteQt uses the generalised modularity matrix
$B^{(g)}$ from Eq.~\eqref{eqn:gen_modularity_mat}, and reruns the above single split pipeline.
Note, that we do not modify the graph after the first community splitting nor the threshold, $\tau$.

If recursion ends before reaching $K$, deteQt forces further splits. Among groups
with an available probe-based bisection, it chooses the group with the smallest
$\chi^2$ $p$-value, without applying the $0.05$ acceptance threshold. If no group
has a probe-based partitioning proposal, a random split happens. Later in the results, we
demonstrate this happens frequently after certain graph sizes, which can be mitigated by 
several proposals.
Since we use the community detection pipeline for further data analysis tasks, we
find it is not necessary to use the zero-overlap deterministic elimination routine
of Ref.~\cite{umeano_quantum_2025}, at least for the problems we choose here. It is
more effective and efficient to grow the probe state dimension, or as we propose,
use more than the leading eigenvector as we do in projeQt
(Section~\ref{ssec:deteqt_extensions}). The restriction to small-overlap and low
Hamming-weight significantly reduces the required circuit depths.
\subsection{projeQt: \texorpdfstring{$k$}{k}-way spectral splitting} \label{ssec:deteqt_extensions}
As graph size increases, recursive bisection accumulates errors at each level because
each split is fixed before the next is made, and it can miss the best division into
more than two groups~\cite{newman2006finding,zhang2015multiway}. Also, since deteQt
operates only on the leading eigenvector, there is no obvious way to produce multiple
communities directly, without recursion. Our second proposed approach is projeQt, a
quantum algorithm which uses multiple leading eigenvectors and can perform
multi-community detection natively.
projeQt acts directly on the generalised modularity matrix $B^{(g)}$ of the group $g$
being split, of $N$ nodes (Section~\ref{sssec:gen_mod_matrix}), padded with zeros to
size $2^n$, $n = \lceil \log_2 N \rceil$, by producing the following state by the
method of
Refs.~\cite{kerenidis2017recommendation,gilyen_quantum_2019,clader2022resources}:
\begin{equation} \label{eq:projeqt_encoding}
	U \ket{0}_a\ket{j}_s = \ket{0}_a \frac{B^{(g)}}{\lambda}\ket{j}_s +
\ket{\perp},
\end{equation}
where $a$ is an ancilla register and $s$ the system register, each of $n$ qubits, and
$\ket{\perp}$ has $a$ outside $\ket{0}$. This state is produced by first loading the
columns, labelled by $j$, into the state
\begin{equation} \label{eq:projeqt_loaders}
	U_R \ket{0}_a\ket{j}_s = \ket{r}_a\ket{j}_s, \qquad \ket{r} \coloneqq \sum_k
\frac{\lVert B^{(g)}_{k\cdot} \rVert}{\lambda}\ket{k},
\end{equation}
with $\lambda = \lVert B^{(g)} \rVert_F$, so that $\ket{r}$ holds the row norms of
$B^{(g)}$. Then, the row index, $i$, is loaded to produce the state
\begin{equation} \label{eq:projeqt_row}
	U_L \ket{0}_a\ket{i}_s = \ket{i}_a\ket{b_i}_s, \qquad \ket{b_i} = \sum_j
\frac{B^{(g)}_{ij}}{\lVert B^{(g)}_{i\cdot} \rVert}\ket{j},
\end{equation}
with $\ket{b_i}$ row $i$ of $B^{(g)}$ normalised. $U = U_L^\dagger U_R$ acts as a form
of block encoding, as post-selecting the ancilla register on $\ket{0}$ gives
\begin{equation} \label{eq:projeqt_block}
	\begin{aligned}
		\bra{0}_a U \ket{0}_a & = \frac{B^{(g)}}{\lambda},
	       \\
		\bra{0,i} U \ket{0,j} & = \braket{i}{r}\braket{b_i}{j} =
\frac{B^{(g)}_{ij}}{\lambda}.
	\end{aligned}
\end{equation}
The circuit performing these operations is composed of $R_y$ and CNOT gates.
After the state of Eq.~\eqref{eq:projeqt_encoding} is prepared, we use the QSVT to
approximate the following function applied to the eigenvalues of $B^{(g)}$:
\begin{equation} \label{eq:projeqt_filter}
	\begin{gathered}
		f(x) = \frac{1}{2}\left[1 + \tanh\!\left(\kappa\,\frac{x - c}{\lVert
B^{(g)} \rVert_F}\right)\right], \\
		c = \max\Bigl(e, \frac{\lambda_{\max}}{2}\Bigr), \qquad e =
\frac{2\lVert B^{(g)}\rVert_F}{\sqrt{N}}.
	\end{gathered}
\end{equation}
This projects out the highest eigenvectors as follows. Write $B^{(g)} = \sum_k
\lambda_k \ket{v_k}\bra{v_k}$, where every $\lvert \lambda_k \rvert \le \lVert B^{(g)}
\rVert_F = \lambda$, so each $\lambda_k/\lambda$ lies in $[-1, 1]$. Applying a QSVT to
the encoding of Eq.~\eqref{eq:projeqt_encoding} transforms each eigenvalue
$\lambda_k/\lambda$ by a polynomial and leaves the eigenvectors unchanged
~\cite{gilyen_quantum_2019}. The circuit applies a polynomial $p$ fitted to $x
\mapsto f(\lambda x)$, so that $p(\lambda_k/\lambda) \approx f(\lambda_k)$. Write $p =
p_{\mathrm e} + p_{\mathrm o}$ for its even and odd parts, and let $U_{\mathrm e}$ and
$U_{\mathrm o}$ be the corresponding QSVT unitaries. The LCU prepares a uniform
superposition over $U_{\mathrm e}$, $U_{\mathrm e}^{\dagger}$, $U_{\mathrm o}$ and
$U_{\mathrm o}^{\dagger}$ on its two-qubit control register $\ell$. Let $a$ be the
block-encoding ancilla register. The adjoint pairs cancel the unwanted imaginary QSVT
blocks; post-selecting the joint all-zero register leaves an operator on the system
register:
\begin{equation} \label{eq:projeqt_up}
	\begin{aligned}
		\ket{0_{\ell a}} & \coloneqq \ket{00}_{\ell}\ket{0}_a,
	       \\
		\mathcal{S}	 & \coloneqq \frac{1}{4}\bigl(U_{\mathrm e}+U_{\mathrm
e}^{\dagger}
		+U_{\mathrm o}+U_{\mathrm o}^{\dagger}\bigr),
	       \\
		\bra{0_{\ell a}} U_p \ket{0_{\ell a}}
				 & = \bra{0}_a\mathcal{S}\ket{0}_a
	       \\
				 & = \frac{1}{2}\left[p_{\mathrm
e}\Bigl(\frac{B^{(g)}}{\lambda}\Bigr)
		+p_{\mathrm o}\Bigl(\frac{B^{(g)}}{\lambda}\Bigr)\right]
	       \\
				 & = \frac{1}{2}p\Bigl(\frac{B^{(g)}}{\lambda}\Bigr)
		= \frac{1}{2}\sum_k p\Bigl(\frac{\lambda_k}{\lambda}\Bigr)
\ket{v_k}\bra{v_k}.
	\end{aligned}
\end{equation}
Since $\frac{1}{2}(1 + \tanh z) = 1/(1 + e^{-2z})$, each eigenvector is weighted by
\begin{equation} \label{eq:projeqt_weight}
	\begin{gathered}
		f(\lambda_k) = \frac{1}{1 + e^{-2\kappa u_k}}, \qquad u_k =
\frac{\lambda_k - c}{\lambda}, \\
		\lim_{\kappa \to \infty} f(\lambda_k) =
		\begin{cases}
			1,	     & \lambda_k > c, \\
			\frac{1}{2}, & \lambda_k = c, \\
			0,	     & \lambda_k < c.
		\end{cases}
	\end{gathered}
\end{equation}
For a fixed eigenvalue away from the cut, convergence to these limits is exponential in
$\kappa$:
\begin{equation} \label{eq:projeqt_tails}
	\begin{aligned}
		1 - f(\lambda_k) & \le e^{-2\kappa u_k},	       & u_k & > 0, \\
		f(\lambda_k)	 & \le e^{-2\kappa \lvert u_k \rvert}, & u_k & < 0,
	\end{aligned}
\end{equation}
The cut in Eq.~\eqref{eq:projeqt_filter} takes the larger of two thresholds. The first,
$e$, sets the spectral scale of the group. Since $\lVert B^{(g)} \rVert_F^2 = \sum_k
	\lambda_k^2$, $\lVert B^{(g)} \rVert_F/\sqrt{N}$ is the root-mean-square
eigenvalue. We
set $e = 2\lVert B^{(g)} \rVert_F/\sqrt{N}$, so that only eigenvalues substantially
larger than this pass the first cutoff. The second, $\lambda_{\max}/2$, prevents a
strongly structured group from retaining every direction above that cutoff: it keeps
only directions whose eigenvalues are comparable to the leading one. All terms in
Eq.~\eqref{eq:projeqt_filter} are accessible by either cheaply computing the Frobenius
norm of $B^{(g)}$, for $e$, or ground-state eigensolving $-B^{(g)}$, for
$\lambda_{\max}$, as deteQt allows. In other words, we do not require explicitly
computing excited states of $-B^{(g)}$. In every result below, we set $\kappa = 32$ and
use the degree-$31$ Chebyshev polynomial
\begin{multline} \label{eq:projeqt_polynomial}
	p_{31}(x) = \sum_{q=0}^{31} a_q T_q(x) \approx f(\lambda x) \\
	= \frac{1}{2}\left[1 + \tanh\left(32\left(x -
\frac{c}{\lambda}\right)\right)\right], \qquad x \in [-1, 1].
\end{multline}
where $T_q$ is the Chebyshev polynomial of degree $q$ and the coefficients $a_q$ are
fitted to this target and bounded by $1$. Future work could use generalized QSP or QSVT
phase-synthesis routines to implement this indefinite-parity polynomial directly,
without the even--odd split and four-branch
LCU~\cite{motlagh2024gqsp,sunderhauf2023gqsvt}.
To see the action of the ideal filter itself, write
\[
	\omega_k \coloneqq \frac{1}{1 +
\exp\left[-2\kappa(\lambda_k-c)/\lambda\right]}.
\]
The corresponding spectral filter and its action on node $i$ are
\begin{equation} \label{eq:projeqt_state}
	\begin{aligned}
		F_\kappa	& = \sum_k \omega_k \ket{v_k}\bra{v_k},
   \\
		F_\kappa\ket{i} & = \sum_k \omega_k \braket{v_k}{i}\ket{v_k},
   \\
		\ket{\phi_i}	& = \frac{F_\kappa\ket{i}}{\lVert F_\kappa\ket{i}
\rVert}.
	\end{aligned}
\end{equation}
The finite polynomial implementation uses the same construction with the filter values
replaced by the corresponding polynomial values.
Each per-node state contains information about how that node participates in the
community structure. The similarity between two nodes therefore indicates whether they
belong in the same communities. We compute this similarity by assembling the Gram
matrix $G$ with components $G_{ij}$:
\begin{equation} \label{eq:projeqt_overlap}
	\begin{aligned}
		G_{ij}			& = \bra{i}F_\kappa^2\ket{j} = \sum_k
\frac{\braket{i}{v_k}\braket{v_k}{j}}{\left[1 +
\exp\left(-2\kappa(\lambda_k-c)/\lambda\right)\right]^2}, \\
		\braket{\phi_i}{\phi_j} & = \frac{G_{ij}}{\sqrt{G_{ii}G_{jj}}}
	\end{aligned}
\end{equation}
In the limit as $\kappa \to \infty$, we have $\omega_k \to 1$ for $\lambda_k > c$ and
$0$ otherwise, so
\begin{equation} \label{eq:projeqt_overlap_limit}
	\begin{aligned}
		\lim_{\kappa \to \infty} G_{ij} & = \sum_{\lambda_k > c}
\braket{i}{v_k}\braket{v_k}{j}, \\
		\braket{\phi_i}{\phi_j} 	& \longrightarrow
		\frac{\displaystyle\sum_{\lambda_k > c} \braket{i}{v_k}\braket{v_k}{j}}
		{\displaystyle\sqrt{\sum_{\lambda_k > c} \lvert\braket{v_k}{i}\rvert^2
				\sum_{\lambda_k > c} \lvert\braket{v_k}{j}\rvert^2}}.
	\end{aligned}
\end{equation}
To estimate the Gram-matrix elements, we use $N(N-1)$ signed Hadamard tests for the
ordered off-diagonal pairs, with the diagonal elements given by the filtered-state
norms (Appendix~\ref{app:kway_algorithm}). The final step is kernel $k$-means with the
number of communities requested for that split.
We accept the resulting partition only if each community has at least $n_{\min} =
	\max\bigl(2,\lfloor N/(2K)\rfloor\bigr)$ nodes, which prevents the split from
producing
singleton or very small communities, and if it raises the modularity of the whole
graph. A group whose largest eigenvalue $\lambda_{\max}$ is not positive is not split
because it has no positive-gain spectral direction.
Since the algorithm can fail, we incorporate several fallbacks into community
selection. First, projeQt can be recursive, initially selecting $K$ communities, then
further splitting the largest group that can be split until the requested $k^*$
communities remain. Unlike deteQt, this is not strictly required, as one can initially
set $k^*=K$ directly. Secondly, if a requested split is not accepted, projeQt reruns
kernel $k$-means with $k=2$ on the same Gram matrix. It then applies
Kernighan--Lin~\cite{kernighan1970efficient}, a separate graph-based refinement, to the
resulting two groups. If no group can be split this way, it takes the two-way labels
returned by kernel $k$-means on the filtered embedding for the group with the largest
modularity gain, even if that gain is negative. It uses a random bisection only when no
group has a filter eigenvalue above its cut (Appendix~\ref{app:settings}).
\subsection{From communities to output} \label{ssec:pipeline_output}
\paragraph{Feature selection} Once the pipeline has partitioned the feature graph
into $K$ communities, one representative feature is selected from each. Within
community $C$, we compute the average distance of each feature to the other
members, with distance $1 - |r_{ij}|$:
\begin{equation} \label{eq:mad}
	\mathrm{MAD}(i) = \frac{1}{|C| - 1} \sum_{\substack{j \in C \\ j \neq i}}
\bigl(1 - |r_{ij}|\bigr).
\end{equation}
The representative is $\arg\min_{i \in C}\, \mathrm{MAD}(i)$, the feature of
	minimum average distance (MAD), which has the largest mean absolute correlation
with
	the rest of $C$ and is the feature most central to its community. This
criterion is
target-agnostic: it does not couple the selected features to a specific response
variable, making it applicable to both supervised and unsupervised settings. For
singleton communities ($|C| = 1$), the sole member is trivially selected. The
	selected features train a random forest classifier on $80\%$ of the samples, and its
	accuracy on the other $20\%$ is reported, averaged over seeds
	(Appendix~\ref{app:settings}).
\paragraph{Clustering} The communities are the clusters. They are scored by their
accuracy under the best
	matching of clusters to labels, and by NMI and ARI.
\section{Results} \label{sec:results}
We evaluate the pipeline on five settings of increasing graph size and complexity.
First, graphs drawn from the SBM of Eq.~\eqref{eqn:sbm} compare both methods with
Louvain (Section~\ref{ssec:results_sbm}). We then apply the pipeline to feature
selection on three real datasets: the Pima Indians Diabetes
dataset~\cite{smith1988using} ($768$ samples, $8$ features augmented to $17$;
Section~\ref{ssec:results_diabetes}), the Wisconsin Breast Cancer
dataset~\cite{street1993nuclear} ($569$ samples, $30$ features;
Section~\ref{ssec:results_breast_cancer}), and MNIST~\cite{lecun1998gradient}
($70{,}000$ samples, $719$ pixel features; Section~\ref{ssec:results_mnist_fs}).
Finally, we apply the same pipeline to unsupervised clustering on MNIST by constructing
a data-point similarity graph rather than a feature correlation graph, scaling to
$2{,}000$ nodes (Section~\ref{ssec:results_mnist_clustering}).
\subsection{Spectral methods vs Louvain on planted partition graphs}
\label{ssec:results_sbm}
\begin{figure}[!htb]
	\centering
	\includegraphics[width=\columnwidth]{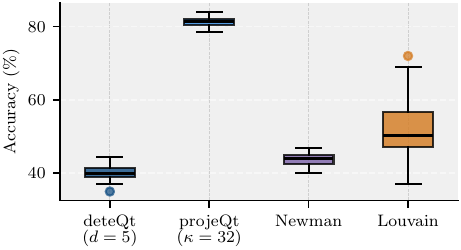}
	\caption{Stability on one SBM graph from Eq.~\eqref{eqn:sbm} with $N = 200$, $K
= 5$, and $r = 3$, across $20$ random seeds.}
	\label{fig_sbm_stability}
\end{figure}

We compare deteQt, projeQt, Newman's leading-eigenvector
method~\cite{newman2006modularity}, and Louvain on graphs drawn from the SBM of
Eq.~\eqref{eqn:sbm}. The ratio $r = p_{\text{in}} / p_{\text{out}}$ controls the
community signal. We use graphs with $N = 200$, average degree $20$, and $K = 4$ or
$10$ planted communities at $r = 1.5$, $3$, and $5$.
\begin{figure}[!t]
	\centering
	\includegraphics[width=\columnwidth]{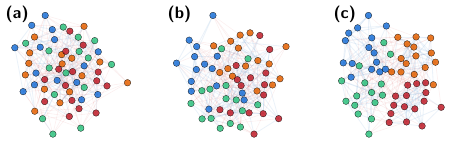}
	\caption{Stochastic block model graphs (Eq.~\eqref{eqn:sbm}, $N = 60$, $K = 4$,
average degree $10$, $r = 1.5$, $3$ and $5$) at increasing community signal strength $r
= p_{\text{in}}/p_{\text{out}}$. Nodes are coloured by planted community, and blue
edges are intra-community, red edges are inter-community. \textbf{(a)}~$r = 1.5$, weak
signal. \textbf{(b)}~$r = 3$, moderate. \textbf{(c)}~$r = 5$, strong.}
	\label{fig_sbm_examples}
\end{figure}
\begin{figure}[!t]
	\centering
	\includegraphics[width=\columnwidth]{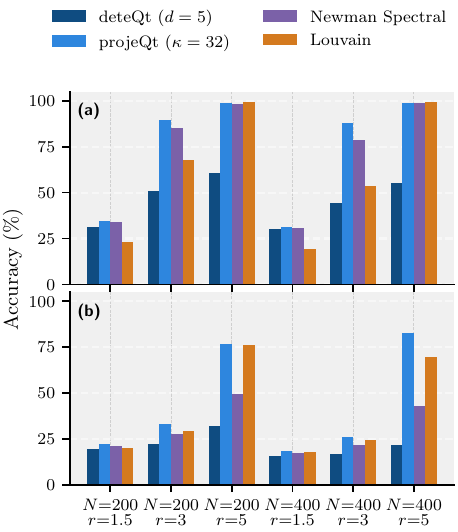}
	\caption{Community recovery accuracy on SBM graphs from Eq.~\eqref{eqn:sbm},
with average degree $20$, averaged over five seeds. \textbf{(a)}~$K = 4$ synthetic
communities. \textbf{(b)}~$K = 10$ synthetic communities. Louvain chooses its own
number of communities and is scored on that partition.}
	\label{fig_sbm_bar}
\end{figure}
Figure~\ref{fig_sbm_bar} reports the fraction of nodes assigned to their planted
community after an optimal one-to-one match. The spectral methods are asked for the
planted $K$, while Louvain automatically chooses its own number of communities. At $r = 1.5$, all
methods are close to chance. At $r = 3$, projeQt exceeds Louvain by $22$ and $34$
percentage points for $K = 4$ at $N = 200$ and $400$, respectively, and by $4$ and $2$
points for $K = 10$. At $r = 5$, projeQt, Newman, and Louvain almost recover all four
communities. For $K = 10$, projeQt and Louvain are level at $N = 200$, while projeQt
leads by $13$ points at $N = 400$. deteQt is below Louvain in eight of the twelve
settings.
We also run each method $20$ times on one SBM graph with $N = 200$, $K = 5$, and $r =
3$. Figure~\ref{fig_sbm_stability} shows average accuracy $81.4\% \pm 1.2\%$ for projeQt,
$57.8\% \pm 0.8\%$ for Newman, $52.4\% \pm 9.0\%$ for Louvain, and $37.6\% \pm 2.2\%$
for deteQt. projeQt's variation across seeds is seven times smaller than Louvain's. Moreover,
Louvain shows a larger variance in the computed results.
\FloatBarrier
\subsection{Diabetes feature graph and selection} \label{ssec:results_diabetes}
\begin{figure}[!htb]
	\centering
\begin{minipage}[b]{0.48\columnwidth}
		\centering
		\includegraphics[width=\textwidth]{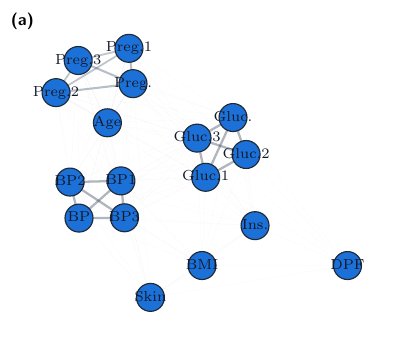}
\end{minipage}
	\hfill
\begin{minipage}[b]{0.48\columnwidth}
		\centering
		\includegraphics[width=\textwidth]{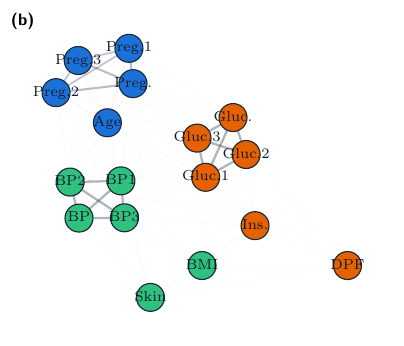}
\end{minipage}
	\caption{Feature correlation graph for the augmented Diabetes dataset
($17$ features) with projeQt community detection. Nodes are features and edges connect pairs
with $|r_{ij}| \geq r_{\mathrm{crit}}$. Text labels identify each source feature and
three synthetic copies, where applicable. \textbf{(a)}~The unpartitioned input graph. \textbf{(b)}~The
three communities returned by projeQt's circuit with $K = 3$. ProjeQt correctly groups the noisy features with their
origin, and since we only specify $K=3$, the non-duplicated features are also included in one of the communities.}
	\label{fig_diabetes_graphs}
\end{figure}

The Pima Indians Diabetes dataset~\cite{smith1988using} contains $768$ samples,
eight clinical features, and a binary label indicating the presence or absence of
diabetes. We augment the dataset with planted features, adding three noisy copies of
Pregnancies, Glucose, and BloodPressure for a total of $17$ features. This provides a 
controlled setting in which the expected community structure is known in advance.
Figure~\ref{fig_diabetes_graphs} shows one circuit run of projeQt with $K = 3$.
Each of the planted (noisy) features can be seen to be grouped together with their origin, 
as expected. The non-duplicated features are assigned to one of the $K=3$ selected communities.
\begin{figure}[t]
	\centering
	\includegraphics[width=\columnwidth]{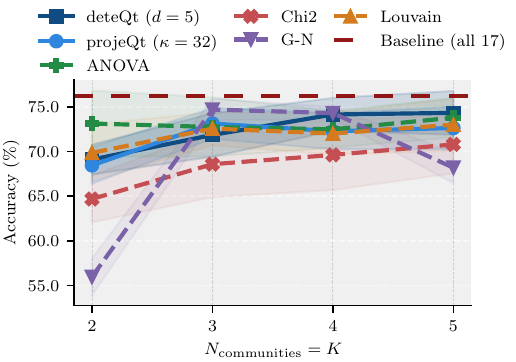}
	\caption{Classification accuracy on the augmented Diabetes dataset as a
function of the number of communities.}
	\label{fig_diabetes_fs}
\end{figure}
For feature selection, we choose one representative feature (or centroid) per community using
MAD and train a random forest classifier on the selected features. We compare deteQt
and projeQt with Louvain, Girvan--Newman~\cite{girvan2002community}, ANOVA, and
chi-squared selection. To compare, we enforce each method to return two to five features. Note, not all community detectors take the selected number of communities as a parameter: when the selected number is not produced in Figure~\ref{fig_diabetes_fs}
we perform a random split or merge. Figure~\ref{fig_diabetes_fs} compares the methods on the $17$-node feature graph, and
we see our proposed methods compete with the state of the art community detectors for this
problem, and sometimes outperforming. Finally, while the selected communities are performative in absolute accuracy, we also
investigate how they affect the model complexity itself (Table\ref{tab:inference_time_diabetes}). Note, while feature selection makes the models faster to fit, the prediction is slower than with all
17 features because the resulting trees grow from $42, 500$ to $51,200 - 92,800$ tree nodes. The larger models required to fit the data likely originates from the fact we are also grouping standalone features with duplicated ones, and only selecting one 
feature per subset. As a result, we are also discarding useful feature information. Nevertheless, among community detection methods, projeQt matches or is faster at both prediction and 
fitting, while deteQt still remains competitive.

\begin{table*}
	\caption{Random forest cost on the augmented Diabetes dataset ($250$ trees, $154$-sample test split), mean $\pm$ std over seeds, in milliseconds of CPU time. Prediction time is the median of $20$ calls per fitted forest, and fitting is timed once. Baseline uses all $17$ features. Average rows report $\bar{t} = \frac{1}{4}\sum_{K=2}^{5} t_K$, with standard deviations propagated assuming independent runs across $K$: $\sigma_{\bar{t}} = \frac{1}{4}\sqrt{\sum_{K=2}^{5}\sigma_K^2}$, where $\sigma_K$ is the standard deviation over seeds at each $K$.}
	\label{tab:inference_time_diabetes}
	\footnotesize
	\setlength{\tabcolsep}{3pt}
	\resizebox{\textwidth}{!}{%
\begin{tabular}{ccccccccc}
	\toprule
	$K$ & Baseline & ANOVA & $\chi^2$ & Mutual info & Girvan--Newman & Louvain & projeQt & deteQt \\
	\midrule
	\multicolumn{9}{c}{\emph{Prediction}} \\
	2 & 11.6 & 12.4 $\pm$ 0.5 & 12.4 $\pm$ 0.4 & 12.6 $\pm$ 0.2 & 13.2 $\pm$ 0.3 & 12.3 $\pm$ 0.5 & \textbf{12.2} $\pm$ 0.3 & 12.4 $\pm$ 0.3 \\
	3 & 11.6 & \textbf{11.9} $\pm$ 0.2 & 12.2 $\pm$ 0.3 & 12.1 $\pm$ 0.3 & 12.0 $\pm$ 0.2 & 12.3 $\pm$ 0.3 & 12.1 $\pm$ 0.3 & 12.4 $\pm$ 0.4 \\
	4 & 11.6 & 12.0 $\pm$ 0.2 & 11.9 $\pm$ 0.1 & 12.4 $\pm$ 0.2 & 12.1 $\pm$ 0.2 & 12.1 $\pm$ 0.4 & 11.9 $\pm$ 0.4 & \textbf{11.9} $\pm$ 0.3 \\
	5 & 11.6 & \textbf{11.7} $\pm$ 0.3 & 12.0 $\pm$ 0.2 & 11.9 $\pm$ 0.2 & 11.8 $\pm$ 0.1 & 11.9 $\pm$ 0.2 & 11.8 $\pm$ 0.4 & 12.0 $\pm$ 0.3 \\
	\midrule
	Average & \textbf{11.600} & \textbf{12.000} $\pm$ 0.162 & 12.125 $\pm$ 0.137 & 12.250 $\pm$ 0.115 & 12.275 $\pm$ 0.106 & 12.150 $\pm$ 0.184 & \textbf{12.000} $\pm$ 0.177 & 12.175 $\pm$ 0.164 \\
	\midrule
	\multicolumn{9}{c}{\emph{Fitting}} \\
	2 & 279 & 212 $\pm$ 6 & 199 $\pm$ 8 & 214 $\pm$ 4 & 217 $\pm$ 5 & 200 $\pm$ 9 & \textbf{198} $\pm$ 5 & 204 $\pm$ 5 \\
	3 & 279 & 204 $\pm$ 3 & 201 $\pm$ 1 & 207 $\pm$ 4 & 202 $\pm$ 3 & \textbf{199} $\pm$ 3 & 201 $\pm$ 4 & 204 $\pm$ 6 \\
	4 & 279 & 237 $\pm$ 5 & 234 $\pm$ 3 & 245 $\pm$ 4 & 238 $\pm$ 3 & 227 $\pm$ 5 & \textbf{220} $\pm$ 6 & 223 $\pm$ 5 \\
	5 & 279 & 227 $\pm$ 4 & 235 $\pm$ 4 & 233 $\pm$ 4 & 223 $\pm$ 3 & 225 $\pm$ 4 & \textbf{217} $\pm$ 6 & 248 $\pm$ 41 \\
	\midrule
	Average & 279.00 & 220.00 $\pm$ 2.32 & 217.25 $\pm$ 2.37 & 224.75 $\pm$ 2.00 & 220.00 $\pm$ 1.80 & 212.75 $\pm$ 2.86 & \textbf{209.00} $\pm$ 2.66 & 219.75 $\pm$ 10.51 \\
	\bottomrule
\end{tabular}%
	}
\end{table*}

%\subsubsection{Effect of thresholding} \label{sssec:graph_construction_results}
%
\subsection{Feature selection on the Breast Cancer dataset}
\label{ssec:results_breast_cancer}

\begin{figure}[t]
	\centering
	\includegraphics[width=\columnwidth]{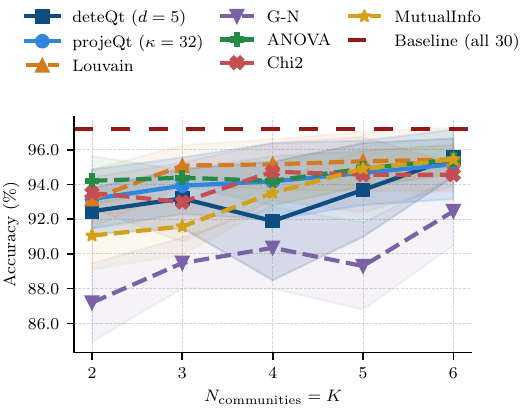}
	\caption{
    Classification accuracy on the Breast Cancer dataset as a function of
    the number of communities. Again, with those where a selected number of communities cannot be selected a-priori, 
    we perform random splitting to make the required number.
    }
	\label{fig_bc_fs}
\end{figure}
Secondly, we test on the Wisconsin Breast Cancer dataset~\cite{street1993nuclear} with no synthetic features.
This dataset contains $569$ samples and
$30$ continuous features computed from digitised images of fine needle aspirates, with
a binary diagnosis label. The $30$ features consist of $10$ base measurements (radius,
texture, etc.) each reported as mean, standard error, and worst value, creating natural
correlation structure among features of the same measurement type.
The feature graph is constructed using all $30$ features with weighted edges and the
statistically derived threshold (Section~\ref{sssec:graph_construction}). As before, we compare deteQt, projeQt, Louvain and Girvan--Newman for community detection, alongside ANOVA, chi-squared, and
mutual
information filter baselines and train a random forest classifier ($500$ estimators) on the selected features. Again, we observe that deteQt and projeQt are competitive in identifying useful features. As with Diabetes, we compare the 
approaches against the baseline model where all $30$ available features are input in the classifier, and community detection methods achieve within $3\%$ of this baseline.
\subsection{Feature selection on MNIST} \label{ssec:results_mnist_fs}
To test the pipeline on a larger feature graph, we apply it to the
MNIST~\cite{lecun1998gradient} image dataset. We treat the $784$ pixels as features and build a
correlation graph. Removing pixels with zero variance leaves $719$ features, which physically means removing the common transparent background from all images. A fully connected graph 
of this size would contain $258, 121$, and using the Pearson correlation matrix (computed
over all $70k$ data samples) prunes this to a dense graph with $162, 474$ edges, $63\%$ of the full graph. We use the statistically derived threshold $r_{\text{crit}} =
0.0074$ for this (Section~\ref{sssec:graph_construction}).

\begin{figure}[t]
	\centering
	\includegraphics[width=\columnwidth]{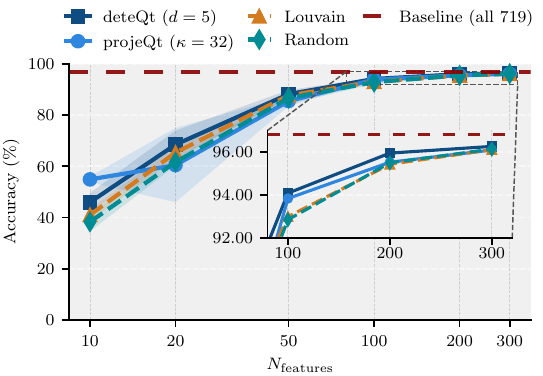}
	\caption{Classification accuracy on MNIST as a function of the number of
selected features ($70, 000$-sample correlation graph, $719$ nodes). Mean and one
standard deviation over five seeds. We again compare to fully features dataset with all $719$ pixels. By $ N_{\text{features}}\approx 200$, we recover within $2\%$ of this maximal performance.}
	\label{fig_mnist_fs}
\end{figure}

Communities in this graph correspond to groups of spatially correlated pixels. As above, we select one representative
pixel per community.
We then train a random forest classifier ($100$ estimators) on the selected pixels, for each community detection method. Figure~\ref{fig_mnist_fs} reports accuracy for projeQt, deteQt (probe-state
bisection, QSVT polynomial $d = 5$), Louvain, and random feature selection. It is shown that at low number of features, the performance of our methods with Louvain and random selection is competitive, and sometimes slightly worse. This is due
to a seed dependence in the results, observable by the error bars indicating one standard deviation. However, above $\sim 50$ nodes, both our methods stabilise and outperform Louvain by a few percent.

\begin{figure}[!htbp]
	\centering
	\includegraphics[width=\columnwidth]{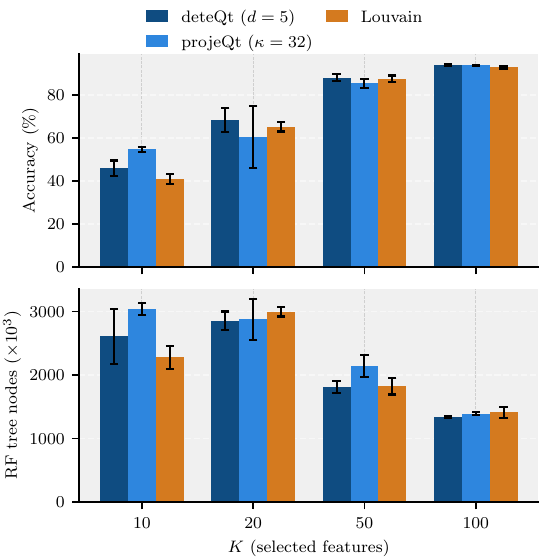}
	\caption{Classification accuracy (left) and random forest model size (right) on
full MNIST ($70{,}000$ samples) as a function of the number of selected features $K$. Results show mean and standard deviation results after 5 independent runs of the pipeline.}
	\label{fig_pareto}
\end{figure}

However, analyzing the methods on MNIST and comparing to the previous two datasets reveals an interesting feature.
Community structure may help more when the redundancy it removes is concentrated
rather than diffuse, as appears to be the case on the $17$- and $30$-node graphs
of Sections~\ref{ssec:results_diabetes} and~\ref{ssec:results_breast_cancer},
but not here. The MNIST experiment therefore highlights a limitation of the
pipeline at larger feature budgets.
From $K = 50$, projeQt's spectral criterion stops supporting further divisions
before $K$ is reached, and a group with no eigenvalue above projeQt's cut is
halved at random: about a third of projeQt's communities at $K = 50$ and two
thirds at $K = 300$ come from such halvings.
Louvain finds about ten communities and reaches every tested budget from $20$
by random halving. deteQt divides further through forced splits taken from the
probe states, which make up between a third and most of its communities from
$K = 200$. At larger budgets, many of projeQt's and Louvain's communities are
random subsets of large groups, so their comparison with random selection
does not isolate the contribution of community structure.
Future work could address this by decoupling the selected number of features
from the number of communities, selecting more than one representative per
community rather than forcing further splits. Random halving could also be
replaced with splits based on within-group diversity or predictive relevance.
Alternative graph constructions and adaptive feature budgets could help
accommodate diffuse redundancy, while controlled experiments could establish
when community structure, representative selection, and forced splits
contribute to downstream performance. Note these are primarily augmentations to the data analysis pipeline itself
rather than the underlying quantum community detection algorithms.
%
%\subsubsection{Model complexity}
%
Figure~\ref{fig_pareto} shows classification accuracy against random forest model size (total number of tree nodes across all $100$ estimators) on the full $70{,}000$-sample MNIST dataset. At $K = 10$ projeQt is the most accurate and has the largest model. From $K = 20$ deteQt has the highest mean accuracy and the smallest mean model, up to $5\%$ smaller than Louvain's, though the size differences, and the accuracy differences at $20$ and $50$ features, are within the seed spread, while projeQt's models are larger than Louvain's at some budgets and smaller at others.

\subsection{Extension to unsupervised clustering} \label{ssec:results_mnist_clustering}
Finally, we demonstrate the application of the quantum community detection pipeline to the problem of unsupervised clustering. As mentioned above, here we replace the feature correlation graph with a data-point similarity graph. To demonstrate this, we apply the
pipeline to direct clustering of the MNIST dataset: each of $N$ images is a node, and a $k$-nearest-neighbour
graph is built in a PCA-reduced space ($n_c = 40$ components, $k = 15$ neighbours,
unweighted edges). We benchmark deteQt, projeQt, and several classical community detection methods on this problem, as above. Here, the number of communities are the number of distinct image classes, $K=10$. The objective is to identify $K=10$ clusters where each cluster $i$ contains the images representing the same digit. In this example, as mentioned above, greedy modularity, Louvain choose their own number of communities but here we do not randomly partition until the specified number of communities are selected. Spectral Newman can take as input a specified number of communities but may fail to produce the required number if no further modularity gains can be made. Instead for these methods, we score them directly on the output partitios as in their standard implementations. For Kernighan--Lin
$10$ classes are produced by bisecting its largest group in turn.

The graph construction here is different to the feature selection case. The resulting kNN graph is
sparse and unweighted ($N$ nodes, resulting in $\sim 7.5N-15N$ edges), whereas the
feature correlation graphs are dense and weighted. The spectral structure is weaker on
sparse graphs, and the distinction between spectral and greedy methods becomes less
clear-cut than on the dense correlation graphs of Sections~\ref{ssec:results_diabetes}
to \ref{ssec:results_mnist_fs}.
\begin{figure}[!htb]
	\centering
	\includegraphics[width=\columnwidth]{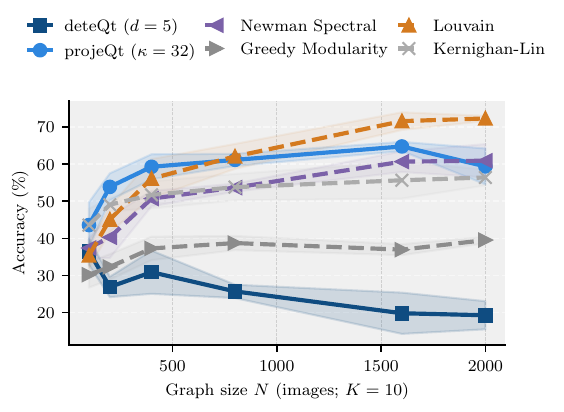}
	\vspace{0.5em}
	\caption{Clustering accuracy on MNIST as a function of graph size $N$ (kNN
similarity graph). Results show average performance after 5 independent runs for each of the tested methods.}
	\label{fig_clustering_accuracy}
\end{figure}

Figure~\ref{fig_clustering_accuracy} shows clustering accuracy as a function of graph size $N$, up to $2000$ graph nodes (executable with $\log_2(N)\approx 10$ qubits, averaged over $5$ seeds. As $N$ scales, deteQt struggles, possibly due to the restriction to the simplest case where $k_{\mathrm{LCU}} = 1$. deteQt decides each split from one eigenvector of a progressively sparser subgraph and needs more splits as $N$ grows.
On these problems, Louvain outperforms our methods, unless we add Louvain-like passes on the community selection output from projeQt and deteQt. This potentially indicates that either feature selection is a more promising problem family, or clustering problems require more dense data points.
\section{Conclusions} \label{sec:conclusions}
In this work we have studied quantum spectral community detection as a tool for data
analysis, applied to feature selection and clustering. We extended deteQt from a single
bipartition to an arbitrary number of communities by recursive bisection and included the generalised modularity matrix. We also
proposed projeQt, which applies a quantum spectral filter to the modularity
matrix. It retains several leading eigenvectors at once and divides a graph into
multiple communities in a single step, without preparing excited states.

Across synthetic and real datasets, both methods are competitive with established
classical community detectors. projeQt recovers planted communities more accurately
than Louvain at moderate signal strength. It also correctly groups redundant features
with their sources, and it performs well on MNIST at small feature budgets and on
clustering graphs of moderate size. deteQt is competitive in feature selection, but it
is weaker on synthetic graphs and on large clustering problems. Neither method
consistently outperforms redundancy-aware feature filters. On the largest graphs, the
greedy optimisation of Louvain still extracts more structure than the spectral criteria
in their current configuration. All results were obtained in classical simulation, so
we make no claim of quantum speedup. Rather, they show that the outputs of these quantum
subroutines are useful inputs to data-analysis pipelines.

Several of the remaining limitations lie in the pipeline rather than in the quantum
algorithms themselves. Future work could decouple the number of selected features from
the number of communities, and replace forced random splits with more informed
criteria. Further directions are a statistically robust acceptance test for deteQt, a
principled choice of projeQt's spectral cut, and more efficient circuit constructions
and readouts. These improvements would benefit classical and quantum spectral detectors
alike.

\bibliographystyle{unsrtnat}
\section*{Declarations}
\paragraph{Funding.} The authors acknowledge support from Fujitsu Research of Europe and Fujitsu Limited.
\paragraph{Competing interests.} The authors declare no competing interests.
\paragraph{Data availability.} All datasets are public benchmarks: the Pima Indians
Diabetes and Wisconsin Breast
Cancer sets from the UCI Machine Learning Repository, and MNIST. 
\paragraph{Code availability.} The code which produced the results in this paper was produced using OpenQARP~\cite{scali2026openqarp}. The OpenQARP repository is available at \url{https://github.com/OpenQARP/openqarp}. We also use NetworkX~\cite{hagberg2008networkx} for graph generation and classical community detectors.
\paragraph{Use of large language models.---}The authors used large language model tools (Claude, Anthropic; Codex, OpenAI) to assist with software development and code review, and editing of the manuscript text and its \LaTeX{} structure. The authors reviewed and verified all outputs and take full responsibility for the content of this work.

\bibliography{main}
\newpage
\appendix
\section{Ablation studies} \label{app:ablations}
\subsection{Threshold influence in graph construction} \label{app:thres_study}

\begin{figure}[!htb]
	\centering
	\includegraphics[width=\columnwidth]{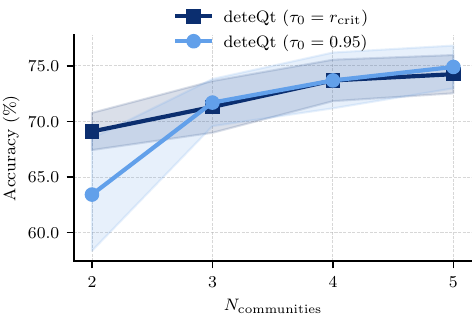}
	\caption{Effect of thresholding on the Diabetes dataset. With $\tau_0 =
r_{\mathrm{crit}}$, rejected splits are retried at higher thresholds. With $\tau_0 =
0.95$, a rejected split is not retried at a higher threshold; forced completion may
still split the group from its probe evidence.}
	\label{fig_threshold_comp}
\end{figure}

First, Figure~\ref{fig_threshold_comp} compares two settings of the edge-weight
thresholding mechanism (Section~\ref{sssec:graph_construction}) on the Diabetes
dataset. Both runs use the same approximate sign function ($d = 5$) and
probe-state bisection. We compare a fixed value $0.95$ against the statistically derived threshold, 
 $r_{\mathrm{crit}}$, which for this instance is $r_{\mathrm{crit}} = 0.071$, keeping $114$ of $136$ edges. The fixed 
threshold in contrast only retains $15$ edges. Due to the increasing threshold rule, the high fixed value of $0.95$
 ensures a failed split is not retried at a higher threshold, while it is allowed when using $r_{\mathrm{crit}}$.
  When $K = 2$ the two perform equivalently in this case. However with $K\geq 3$, $\tau_0 = r_{\mathrm{crit}}$ outperforms.
   All experiments in the paper use the critical threshold unless specified.

\subsection{Low degree QSVT in deteQt}\label{ssec:qsvt_sign_approx}
\begin{figure}[!htbp]
\centering
\includegraphics[width=\columnwidth]{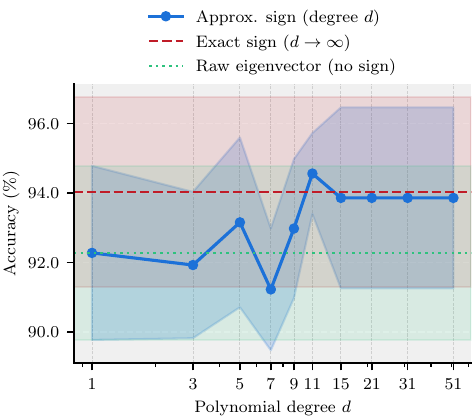}
\caption{Breast Cancer classification accuracy at $K=4$ for the raw
eigenvector, the degree-$d$ QSVT sign polynomial, and exact signs in deteQt. Up to
$d=21$ we evaluate by directly simulating the circuits, reverting to classical emulation of the
effective polynomial action for higher degrees. Shading
shows one standard deviation over five seeds.}
\label{fig_sign_sweep}
\end{figure}

Next, Figure~\ref{fig_sign_sweep} compares deteQt's classification accuracy on the Breast
Cancer dataset in two scenarios. Firstly, when its amplitudes are rounded
with the degree-$d$ QSVT polynomial, and secondly when they are rounded to exact signs. The `raw'
eigenvector is equivalent to the degree-1 odd polynomial up to rescaling (or, using the output of the
ground state preparation circuit directly and bypassing the sign refinement). We observe that as $d$ increases,
the approximate sign via the QSVT approaches the accuracy of the exact sign rounding, which in turn outperforms the
raw sign directly. The approximate saturation occurs by $d=15$ so only moderate-depth circuits are required for this
problem.
\section{Baseline methods} \label{app:baselines}
The paper compares several classical and canonical community-detection and feature-selection
methods, which we describe briefly here.
\subsection{Classical community detection baselines}
\paragraph{Girvan--Newman.} Girvan--Newman~\cite{girvan2002community}
repeatedly removes an edge with the highest edge \emph{betweenness}, splitting the graph into
communities, selecting the edge with largest betweenness
$b(e)=\sum_{s<t}\sigma_{st}(e)/\sigma_{st}$, where $\sigma_{st}$ counts
shortest paths from $s$ to $t$ and $\sigma_{st}(e)$ counts those using $e$. For
feature-selection comparisons, we take the partition after $K-1$ splits.
\paragraph{Greedy modularity.} The Clauset--Newman--Moore method
\cite{clauset2004} starts with one community per node and repeatedly merges the pair
that gives the largest modularity increase $\Delta Q_{ab}=Q(\mathcal{P}_{a\cup b})-
Q(\mathcal{P})$, where $Q(\mathcal{P})=\sum_{C\in\mathcal{P}}
[l_C/m-(d_C/2m)^2]$, $l_C$ is the internal edge weight and $d_C$ the sum of degrees
in $C$. It chooses its own community count and is scored on that partition in the
clustering results.
\paragraph{Kernighan--Lin.} Kernighan--Lin~\cite{kernighan1970efficient}
refines a bipartition by moving nodes between its two parts to reduce the cut
$C(A,B)=\sum_{i\in A,j\in B}w_{ij}$. As a standalone baseline, the largest current group is bisected repeatedly until $K$ groups
are obtained. We use Kernighan--Lin as a refinement step in projeQt of a proposed
bisection.
\paragraph{Newman spectral.} Newman's method~\cite{newman2006finding}
recursively bisects groups using $s_i=\operatorname{sign}(v_{\max,i})$, where
$B^{(g)}v_{\text{max}}=\lambda_{\text{max}}v_{\text{max}}$, where $v_{\text{max}}$ is the leading eigenvector, and then applies vertex-move refinement. The proposed split's
modularity gain is $\Delta Q=s^{\mathsf T}B^{(g)}s/(4m)$. In fixed-$K$
comparisons, it continues splitting while an accepted spectral split is available. If
it cannot reach $K$, it uses seeded random bisection. If it produces more than $K$
groups, it merges the smallest groups. The clustering comparison can instead stop at
the native partition when no further split is accepted.
\subsection{Feature selection baselines}
\paragraph{ANOVA $F$-test.} The one-way ANOVA $F$ statistic ranks features by
the ratio of between-class to within-class variation. It evaluates each feature
independently, with
$F_j=[SS_{B,j}/(C-1)]/[SS_{W,j}/(n-C)]$, where $SS_{B,j}$ and $SS_{W,j}$ are the
between-class and within-class sums of squares for feature $j$.
\paragraph{Chi-squared test.} The chi-squared score measures dependence
between a feature's binned values and the class label, using
$\chi_j^2=\sum_{b,c}(O_{bc}-E_{bc})^2/E_{bc}$ for observed and expected bin-label
counts. Features are ranked by this score.
The inputs are shifted to be non-negative, as required by the test.
\paragraph{Mutual information.} Mutual information ranks each feature by
its estimated dependence on the target,
$I(X_j;Y)=\sum_{x,y}p(x,y)\log[p(x,y)/(p(x)p(y))]$, using the nearest-neighbour
estimator in scikit-learn.
\paragraph{mRMR.} Minimum-redundancy maximum-relevance first selects the
feature with largest $I(X_j;Y)$, then adds the feature
$j^*=\arg\max_{j\notin S}[I(X_j;Y)-|S|^{-1}\sum_{s\in S}I(X_j;X_s)]$, where $S$
is the selected set~\cite{peng2005feature}.
\clearpage
\section{Unified pipeline} \label{app:pipeline}
\begin{figure}[!htb]
	\centering
	\resizebox{0.82\columnwidth}{!}{%
	\begin{tikzpicture}[
		box/.style={draw=black!55, rounded corners=2pt, fill=black!3,
			text width=3.15cm, align=center, inner sep=5pt, font=\scriptsize},
		circuit/.style={box, draw=blue!60!black, fill=blue!5},
		readout/.style={box, draw=orange!75!black, fill=orange!7},
		arrow/.style={-{Stealth[length=4pt]}, draw=black!70, line width=0.6pt}]
		\node (input) [box] {Input graph $\mathcal{G}$};
		\node (matrix) [box, below=0.32cm of input] {Modularity matrix $B$};
		\node (det) [circuit, below left=0.48cm and 0.22cm of matrix] {
			\textbf{deteQt circuit}\\
			VQE $\to$ diagonal amplitude encoding $\to$ sign QSVT $\to$ probe LCU};
		\node (detout) [readout, below=0.22cm of det] {
			$S$ probe outcomes over $N$ probes $\to$ $\chi^2$-gated bisection. Recurse to produce $K$ communities};
		\node (pro) [circuit, below right=0.48cm and 0.22cm of matrix] {
			\textbf{projeQt circuit}\\
			Block-encode $B^{(g)}$ $\to$ QSVT filter $f_\kappa$ $\to$ filtered node states};
		\node (proout) [readout, below=0.22cm of pro] {
			State overlaps $\to$ kernel $k$-means into $K$ communities};
		\node (result) [box, below=0.4cm of $(detout.south)!0.5!(proout.south)$]
			{$K$ communities};
		\draw [arrow] (input) -- (matrix);
		\draw [arrow] (matrix.south) -- ++(0,-0.16) -| (det.north);
		\draw [arrow] (matrix.south) -- ++(0,-0.16) -| (pro.north);
		\draw [arrow] (det) -- (detout);
		\draw [arrow] (pro) -- (proout);
		\draw [arrow] (detout.south) |- (result.west);
		\draw [arrow] (proout.south) |- (result.east);
	\end{tikzpicture}%
	}
	\caption{Unified pipeline of community detection for data analysis. Blue boxes show the circuit stages and orange boxes show classical post-processing. deteQt measures probe outcomes and selects a bisection, recursing until the selected number of communities are found. projeQt
measures state overlaps and applies kernel $k$-means to directly output a selected number of communities, optionally recursively.}
	\label{fig:flow_unified}
\end{figure}
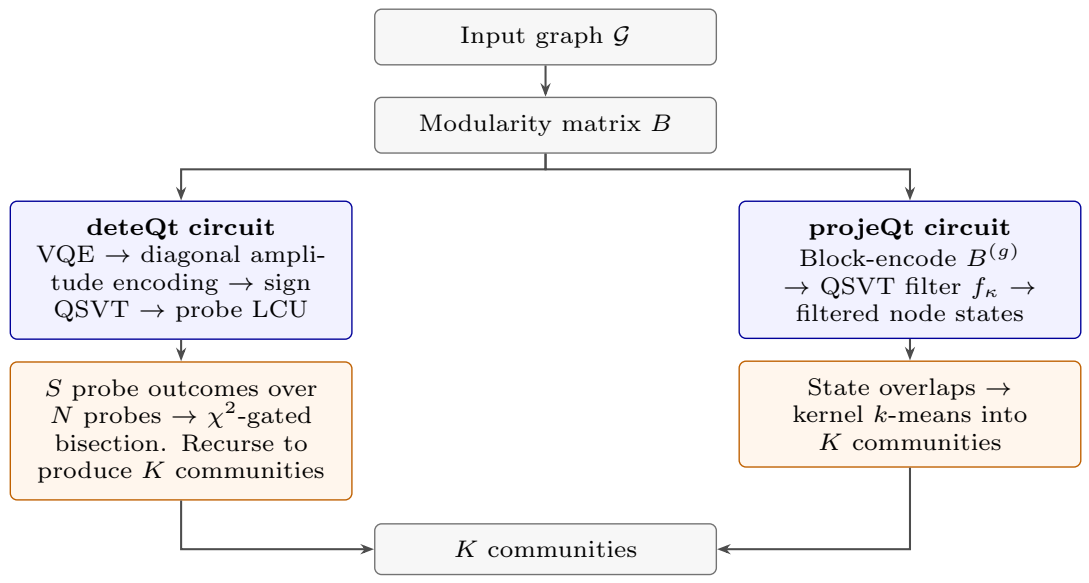
\begin{figure}[!htb]
\centering
\includegraphics[width=0.85\columnwidth]{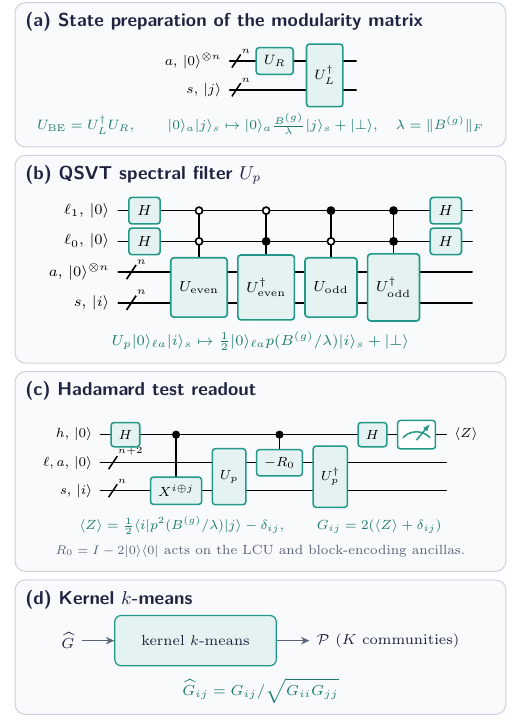}
\caption{\textbf{Circuit ingredients in projeQt algorithm.} \textbf{a)} Circuit to prepare rows and columns of the modularity matrix. \textbf{b)} QSVT implementing the spectral thresholding filter, with odd and even parities implemented via an LCU. \textbf{c)} Hadamard-test readout of Gram-matrix elements $G_{ij}$. \textbf{d)} Kernel $k$-means applied to the Gram matrix $G$ to resolve $K$ communities.}
\label{fig:app_projeqt_circuit}
\end{figure}
\begin{figure}[!htb]
\centering
\includegraphics[width=0.96\columnwidth]{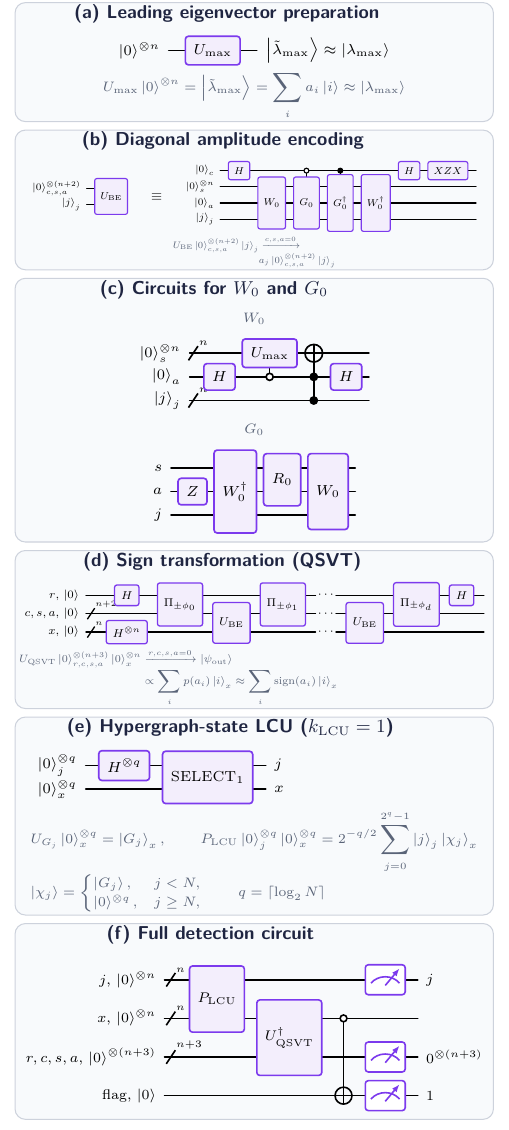}
\caption{\textbf{Circuit ingredients for the deteQt algorithm~\cite{umeano_quantum_2025},
restricted to $k_{\mathrm{LCU}}=1$.} \textbf{a)} Preparation of the leading
 eigenvector $\ket{\lambda_{\max}}$, for example by VQE. \textbf{b)} Block encoding
that places the amplitudes of $\ket{\lambda_{\max}}$ on the diagonal. \textbf{c)}
Subcircuits $W_0$ and $G_0$ used in this block encoding (see Refs.~
\cite{rattew_non-linear_2024,umeano_quantum_2025} for full details). \textbf{d)}
QSVT approximation of the \texttt{sign} function applied to the amplitudes of
$\ket{\lambda_{\max}}$. \textbf{e)} LCU preparation of the $k_{\mathrm{LCU}}=1$
hypergraphs (Hamming weight one). \textbf{f)} Full detection circuit computing
overlaps between the signed eigenvector and candidate probe states.}
\label{fig:app_deteqt_circuit}
\end{figure}
\begin{figure*}[tp]
\begin{algorithm}
\caption{projeQt community retrieval from an input graph}
\label{alg:projeqt_split}
\label{app:kway_algorithm}
\small
\begin{algorithmic}[1]
\Require graph $\mathcal{G}$ and counts $2\le K\le k^*\le |V(\mathcal{G})|$
\Ensure partition $\mathcal{P}$ of $V(\mathcal{G})$ into $k^*$ communities
\State $N_0 \gets |V(\mathcal{G})|$, $n_{\min} \gets \max(2,\lfloor N_0/(2k^*)\rfloor)$
\State $\mathcal{P} \gets \{V(\mathcal{G})\}$
\State $\mathcal{P}' \gets \Call{projeQt}{\mathcal{G},V(\mathcal{G}),K,n_{\min}}$ \Comment{request $K$ communities}
\If{$\mathcal{P}'$ is accepted}
\State $\mathcal{P} \gets \mathcal{P}'$
\EndIf
\While{$|\mathcal{P}|<k^*$}
\State accepted $\gets$ false
\For{communities $g \in \mathcal{P}$ in decreasing size with $|g|>\max(3,n_{\min})$}
\State $R \gets k^* - |\mathcal{P}| + 1$
\If{$R \ge 3$}
\State $K_g \gets \min(R,5)$
\Else
\State $K_g \gets 2$
\EndIf
\State $\mathcal{P}' \gets \Call{projeQt}{\mathcal{G},g,K_g,n_{\min}}$
\If{$\mathcal{P}'$ is accepted}
\State $\mathcal{P} \gets (\mathcal{P} \setminus \{g\}) \cup \mathcal{P}'$
\State accepted $\gets$ true
\State \textbf{break}
\EndIf
\EndFor
\If{not accepted}
\State form a bisection by two-way kernel $k$-means on $\widehat G$ for each $g$ with $\lambda_{\max}(B^{(g)})>c$
\If{at least one proposal exists}
\State choose the highest-$\Delta Q$ proposal, even if negative
\Else
\State bisect the largest community at random
\EndIf
\State replace the selected community by this bisection
\EndIf
\EndWhile
\State \Return $\mathcal{P}$
\end{algorithmic}
\end{algorithm}
\vspace{6pt}
\begin{algorithm}
\caption{projeQt split proposal for one community}
\label{alg:projeqt_trysplit}
\small
\begin{algorithmic}[1]
\Procedure{projeQt}{$\mathcal{G},g,K_g,n_{\min}$}
\Require graph $\mathcal{G}$, community $g$, requested count $K_g$, minimum size $n_{\min}$
\Ensure accepted partition of $g$, or failure
\State form $B^{(g)}$ as in Eq.~\eqref{eqn:gen_modularity_mat}
\State estimate $\lambda_{\max}(B^{(g)})$ by VQE for the ground state of $-B^{(g)}$, as in deteQt
\State $\lambda \gets \lVert B^{(g)}\rVert_F$, $e \gets 2\lambda/\sqrt{|g|}$
\State $c \gets \max(e,\lambda_{\max}(B^{(g)})/2)$ as in Eq.~\eqref{eq:projeqt_filter}
\State choose degree-$d$ polynomial $p_d$ approximating $f$ in Eq.~\eqref{eq:projeqt_filter}
\If{$\lambda_{\max}(B^{(g)}) > 10^{-12}$ and $|g| > 2$}
\State construct $U_{\mathrm{BE}}=U_L^\dagger U_R$ for $B^{(g)}/\lambda$ from the row and column loaders in Eq.~\eqref{eq:projeqt_loaders}~\cite{kerenidis2017recommendation,gilyen_quantum_2019,clader2022resources}
\For{each node $i$ in $g$}
\State prepare and normalise $p_d(B^{(g)}/\lambda)\ket{i}$ as $\ket{\phi_i}$
\EndFor
\State form $\widehat G_{ij}=\braket{\phi_i}{\phi_j}$
\If{$|g| \ge 2K_g$} \Comment{necessary for two nodes per community; $n_{\min}$ is enforced below}
\State $\mathcal{P}' \gets$ kernel $k$-means on $\widehat G$ with $K_g$ communities
\If{every community $C\in\mathcal{P}'$ has $|C|\ge n_{\min}$ and $\Delta Q(\mathcal{P}')>0$}
\State \Return $\mathcal{P}'$
\EndIf
\EndIf
\If{$|g|>2$}
\State $\mathcal{P}_2 \gets$ two-way kernel $k$-means on $\widehat G$
\State refine by Kernighan--Lin
\If{both communities meet $n_{\min}$ and $\Delta Q(\mathcal{P}_2)>0$}
\State \Return $\mathcal{P}_2$
\EndIf
\EndIf
\EndIf
\State \Return failure
\EndProcedure
\end{algorithmic}
\end{algorithm}
\end{figure*}
\clearpage
For completeness, Algorithm~\ref{alg:projeqt_split} gives the full projeQt algorithm
for community retrieval from the input graph. It shows one recursive implementation:
projeQt first requests $K$ communities, then repeatedly splits a current community
until it reaches the target $k^*$. Recursion is optional; as discussed in the main
text, the initial split can request $k^*$ communities directly. The circuit
ingredients are shown in Fig.~\ref{fig:app_projeqt_circuit}. Algorithm~\ref{alg:projeqt_trysplit}
shows a single instance of the projeQt quantum community detector proposing
communities via spectral thresholding, overlap estimation, and classical refinement.
	\section{Experiment hyperparameters} \label{app:settings}
	Table~\ref{tab:settings} gives the settings behind every result in
	Section~\ref{sec:results}.
\begin{table}[!htb]
		\caption{Run settings. $N_0$ is the size of the whole graph and $G$ the
current number of groups.}
		\label{tab:settings}
		\small
		\centering
		\begin{tabular}{p{0.37\columnwidth}p{0.59\columnwidth}}
			\toprule
			setting 				& value
				       \\
			\midrule
			\multicolumn{2}{l}{\textit{deteQt}}
				       \\
			probe state measurements $S$ 			& $1000$
				       \\
			 $k_{\mathrm{LCU}}$ & $1$
				       \\
			sign polynomial 			& odd, degree $5$,
fitted to $\tanh(10x)$			    \\
			split gate				& $\chi^2$ on $S p_j$
against uniform, $p \le 0.05$		 \\
			$\tau_0$, step, ceiling	& $r_{\mathrm{crit}}$
(unit-weight graphs $0$), $0.1$, $1.0$	 \\
			split order				& first in, first out
				       \\
			VQE depth				& $L=8-32$ layers of HEA ansatz \\
			\midrule
			\multicolumn{2}{l}{\textit{projeQt}}
				       \\
			sharpness $\kappa$			& $32$,
                    \\
			polynomial degree			& $31$
				       \\
			cut $c$ 				& $\max(2\lVert
B^{(g)}\rVert_F/\sqrt{N}, \lambda_{\max}/2)$   \\
			kernel $k$-means			& $k$-means++ starts,
$20$ restarts				 \\
			parts per split 			& $K$ first, then
$\min(K - G + 1, 5)$ if $\ge 3$, else $2$    \\
			minimum part size				& $\max(2, \lfloor
N_0/2K \rfloor)$ nodes			    \\
			Kernighan--Lin				& up to $20$ rounds
				       \\
			\midrule
			\multicolumn{2}{l}{\textit{evaluation}}
				       \\
			random forest				& $250$ trees
(Diabetes), $500$ (Breast Cancer), $100$ (MNIST) \\
			train/test split			& $80/20$, not
stratified					\\
			\bottomrule
		\end{tabular}
	\end{table}

\end{document}